# Systematic Target Function Annotation of Human Transcription Factors

Yong Fuga Li[1, 2], Russ B. Altman[2, 3]

Affiliations:

[1]Stanford Genome Technology Center, Stanford, California, USA;

[2]Department of Bioengineering, Stanford University, Stanford, California, USA;

[3]Department of Genetics, Stanford University, Stanford, California, USA

**Correspondence:**

Russ B. Altman (russ.altman@stanford.edu)

**Running Title:** Function Annotation of Transcription Factors

# Abstract

Transcription factors (TFs), the key players in transcriptional regulation, have attracted great experimental attention, yet the functions of most human TFs remain poorly understood. Recent capabilities in genome-wide protein binding profiling have stimulated systematic studies of the hierarchical organization of human gene regulatory network and DNA-binding specificity of TFs, shedding light on combinatorial gene regulation. We show here that these data also enable a systematic annotation of the biological functions and functional diversity of TFs. We compiled a human gene regulatory network for 384 TFs covering the 146,096 TF-target gene relationships, extracted from over 850 ChIP-seq experiments as well as the literature. By integrating this network of TF-TF and TF-target gene relationships with 3,715 functional concepts from six sources of gene function annotations, we obtained over 9,000 confident functional annotations for 279 TFs. We observe extensive connectivity between transcription factors and Mendelian diseases, GWAS phenotypes, and pharmacogenetic pathways. Further, we show that transcription factors link apparently unrelated functions, even when the two functions do not share common genes. Finally, we analyze the pleiotropic functions of TFs and suggest that increased number of upstream regulators contributes to the functional pleiotropy of TFs. Our computational approach is complementary to focused experimental studies on TF functions, and the resulting knowledge can guide experimental design for discovering the unknown roles of TFs in human disease and drug response.

# Introduction

Regulation of gene expression is essential for the realization of cell type specific phenotypes (Whyte et al. 2013) during normal development (Reik 2007) and the adaptation of cellular organisms to their environment (López-Maury et al. 2008). To a large degree, transcriptional regulation occurs through the interaction of protein factors with the genomic DNA (Lenhard et al.

2012). Multiple proteins, including the chromatin remodelers, transcription factors, cofactors and other transcription initiation factors (Perissi and Rosenfeld 2005) work in coordination to regulate the spatial-temporal details of gene expression. Transcription factors (TFs) bind DNA in a sequence specific manner and mediate the integrations of other proteins with specific target genes for fine-granular expression control (Maniatis et al. 1987).

The pivotal role of TFs in development and cell identity determination is highlighted by the induced pluripotent stem cell (iPSC) technology (Takahashi and Yamanaka 2006; Park et al. 2008) and trans-induction techniques (Lee and Young 2013; Jopling et al. 2011), in which the introduction of just a few specific transcription factors is sufficient for converting fibroblast cell into pluripotent stem cell, or converting one differentiated cell type, e.g. pancreatic exocrine cell, directly into another differentiated cell type, e.g. β-cell. In addition, transcription factors are key players controlling diverse physiological functions ranging from metabolism (Yamashita et al. 2001; Wang et al. 2011), chemical and mechanical stress responses (Kaspar et al. 2009; Tothova et al. 2007; Kumar and Boriek 2003; Mendez and Janmey 2012), song-learning (Whitney et al. 2014; Pfenning et al. 2012), to longevity and aging (Greer and Brunet 2005; Salih and Brunet 2008; Tilstra et al. 2012). Many TFs are directly involved in diseases, such as cancer, diabetes, and neural developmental disorders (Lee and Young 2013).

Transcription factors have attracted intense research attention (Yusuf et al. 2012); yet the biological functions of most TFs are still poorly understood. The number of human TFs is estimated to be around 1500-2000 based on DNA-binding domain evidence (Vaquerizas et al. 2009; Ravasi et al. 2010; Ashburner et al. 2000; Kummerfeld and Teichmann 2006). In total, the sequence-specific DNA-binding activities of only 564 TFs are confirmed by experimental evidence and the existence of an additional 490 is supported indirectly by phylogenetic evidence or author claims (The Gene Ontology Consortium 2015). Limited knowledge is available on the biological functions of most TFs, with a small number of "famous" TFs such as TP53 attracting much attention (Vaquerizas et al. 2009). However, recent development of high-throughput technology such as ChIP-seq and DNase-seq (Furey 2012) provide an unprecedented amount of data on gene regulation, with binding profiles for over 100 transcription factors from ENCODE alone (Gerstein et al. 2012). This has spurred systematic data-driven studies on transcriptional regulation, such as the discovery of cis-regulatory motifs (Kheradpour and Kellis 2014; Jolma et al. 2013), the mapping of the hierarchical architecture of human gene regulatory network, and the modeling of combinatorial regulation (Gerstein et al. 2012; Cheng et al. 2011b; Neph et al. 2012; Bernstein et al. 2012; Jiang and Singh 2014). At the same time, analytics tools have been developed for annotating ChIP-seq data (Ji et al. 2011; Zhang et al. 2008; Spyrou et al. 2009), some allowing analysis of gene ontology term enrichment for the binding sites (Zhu et al. 2010; McLean et al. 2010; Welch et al. 2014).

In this study, we integrate the existing knowledge about functions and phenotypes of human genes with the transcriptional regulatory network to study the functions of human transcription factors. We define the "target functions" of a TF as the statistical overrepresented functions among its target genes, and provide a systematic annotation of TF functions, ranging from metabolic pathways to disease phenotypes. In parallel, we define the functional similarity of two-TFs based on their target gene overlaps, independent of the availability of gene function annotations, and annotate each TF by functionally similar TFs (**Fig. 1**). We study the pleiotropic functions of individual TFs and show that multifunctionality is associated with the number of upstream regulators of the TFs. With these analyses, we demonstrate a computational approach for achieving systematic understanding of transcription factor functions.

# Results

## The Compendium of Human Transcription Factor Target Genes

We compiled a Transcription Factor-Target Gene (TFTG) data compendium covering the direct transcriptional regulation targets of 384 unique transcription factors (TFs) extracted from over 850 ChIP-seq experiments as well as the literature with low throughput experimental evidence. Low throughput experiments, ENCODE ChIP-seq, and other sources of ChIP-seq data are complementary in their TF coverage. 149 (39%) TFs are covered only by high-throughput experiments, among which 52 (35%) are covered by the ENCODE consortium (Gerstein et al. 2012; Bernstein et al. 2012), 107 are covered by individual research labs (based on data published by October 2013). Meanwhile 122 (32%) TFs are only from low throughput experiments, and 113 (29%) TFs are from both low and high throughput experiments (**Fig. S1A**).

There are in total 16,967 unique target genes (TGs) of TFs, including both transcription factors and non-transcription factors. We filtered the target genes identified in high-throughput experiments to achieve an estimated false discovery rate of 0.01. Combining all sources, 146,096 TF-TG relationships are obtained. Each gene is regulated by 8.6 TFs on average, while each TF regulates 380.5 genes (see **Fig. S1B**). 63% of target genes are each regulated by 5 or more TFs, while 18% are each regulated by a single TF. Most TFs also have regulators within the compendium, with the exception of 14 TFs that appear to be master regulators, including BCOR, GLI2, HLF, HNF4G, MAZ, NELFE, NFATC1, NOTCH1, PHOX2A, RXRA, STAT4, SOX10, TEAD2 and THRA, although RXRA and SOX10 are self-regulated.

## Defining the Target Functions of Transcription Factors

Transcription factors perform their functions by i) interacting with proteins and cis-regulatory elements and ii) consequently regulating the expression of downstream target genes. There are hence two aspects of functions for a TF, the molecular functions of a TF that enables its regulation of the target genes, and the biological functions exerted by the genes that are under control of the TF. Formally, we define the *target functions* (e.g. target diseases, target signaling pathways) of a transcription factor as the consensus functions of the target genes, and we identify the target functions of a TF by detecting the enrichment of functional terms in the target genes. The target genes as a whole precisely define the biological functions regulated by a TF, while the target functions summarize the functional impacts upon perturbation of a transcription factor.

We first compiled 3715 functional concepts covering molecular to organism level functions (**Table S1)**, including Mendelian diseases from OMIM, disease and phenotype associations from dbGAP genome wide association studies (GWAS), pharmacokinetic (PK) and pharmacodynamics (PD) pathways from PharmGKB, signaling and metabolic pathways from Reactome, and molecular functions and biological processes from Gene Ontology (GO). There are significant overlaps among the genes annotated in the six sources, yet each source has some unique genes (**Fig. S2.**)

We then confirmed the presence of functional signals in the TFTG compendium, i.e., TFs are not randomly targeting functionally unrelated genes, and the TFTG compendium contains significant number of true target genes. We compared the TF-Function associations obtained using real TFTG compendium against that obtained using a randomized compendium, where we constructed the fake TFs to have the same number of random target genes as the corresponding real TFs. We observed 237,566 TF-Function pairs with p-values for real TFs smaller than the corresponding p-values for the fake TFs, compared to 155,801 pairs showing the opposite relationship (**Fig. 2**). To estimate the total number of true associations present for the real TFs, we assume i) true associations for real TFs are all in the upper triangle, i.e. having p-values from real compendium less than corresponding p-values from randomized compendium; ii) false associations for real TFs equally distribute in the upper and lower triangle, i.e. having similar p-values from the real and fake TFs. This leads to an estimated 81,765 true target function annotations for the real TFs. The ratio between the true and false associations is larger at the smaller p-value regions (**Fig. 2** inset). At p-value cutoff of 0.001, there are 16,158 associations for real TFs and 999 for fake TFs, corresponding to an FDR of 6.18%; while at p-value cutoff of 0.0001, there are 9,132 associations for real TFs but only 130 for fake TFs, corresponding to an FDR of 1.42%.

## Gene Universe Impacts the Detection of Target Functions

The target functions of a transcription factor are detected by identifying statistically significant enrichment of functional terms among the target genes of the TF. A critical step for obtaining proper statistics for enrichment analysis is the choice of gene universe, which is the "allowed" set of genes that restrict the target genes of a TF as well as the member genes of a functional term to be used in determining statistical associations. In **Fig. S3**, we show the example of TF SP1 and functional term "immune system". The choice of gene universe affects not only the significance (p-value) but also the direction of TF-target function association. In general, an overly large gene universe inflates the strength of positive association, i.e. enrichment of functional terms, while an overly restrictive gene universe inflates the strength of negative association, i.e. depletion of functional terms.

We suggest that the gene universe must be chosen based on the implicit limitations of each type of functional annotations stemming from how the annotation was obtained, thus generally providing a smaller and hence more conservative universe. For manual curation, such as OMIM and PharmGKB, the function annotations are limited by the available literature. We hence constructed a conservative "Literature Rich" gene universe that includes protein-coding genes annotated by one or more sources from OMIM, PharmGKB, GO BP, GO MF, Reactome, KEGG, and Biocarta. For machine annotations coming from high throughput experiments followed by computational filtering, such as the GWAS phenotype annotations, we use the "Coding Genes" as a conservative universe (see Methods for more details). We disregard non-coding genes as they are generally poorly annotated. We used the Literature Rich gene universe to detect target Mendelian diseases, pharmacogenomic pathways, signaling/metabolic pathways, molecular functions, and biological processes, and used the Coding gene universe to detect target phenotypes studied in GWAS.

## Transcription Factor-Target Function Network

At false discovery rate of 0.05, we identified 9,747 significant TF-target function relationships using the conservative gene universes (**Fig. 3A**). The TF-target function associations form a scale-free network (Barabási 1999), with power law distributions for both the number of target functions per TF and the number of TFs per target function (**Fig. 3B** and **Fig. S4A**). Overall, 279 (73%) transcription factors are annotated by at least one functional term (Supplemental Material Section 1.1). The lack of the annotations of the remaining TFs is likely due to small sample size, i.e. number of target genes. The un-annotated TFs have 26.3 target genes on average, compared to 519.0 target genes on average for annotated TFs (**Fig. S4B**). An average TF is positively associated with 0.47 Mendelian diseases, 0.052 GWAS phenotypes or diseases, 0.26 pharmacogenomic pathways, 11.2 signaling and metabolic pathways, 7.9 biological processes, and 1.4 molecular functions (**Table S2**). Extensive regulator sharing is observed among different types of gene functions (**Fig. S5A**), while we also observed biases of 62 TFs towards specific types of functions (**Fig. S5B** and Supplemental Material Section 1.2).

## Target Functions Predict Known Functions of Transcription Factors

We globally validate the TF-target function relationships by comparing them against the known functions of these transcription factors. Of course, our TF-target function relationships do not necessarily map to a TF-function relationship that is covered by existing gene annotation databases. For example, AHR targets molecular function “oxygen binding”, indicating that AHR regulates proteins that bind oxygen and likely catalyze oxidation reactions, but this does not mean oxygen binding is a molecular function of AHR protein itself. HNF1A targets many pharmacokinetic pathways (**Fig. 3G**), but HNF1A is naturally not an annotated member of these pharmacokinetic (PK) pathways, as the PK pathways in PharmGKB focus on the metabolic enzymes and transporters of drugs. Despite that, we find that the TF-target function associations can predict the known TF-target function relationships for all six types of functions. An overall area under the ROC curve (AUC) of 0.80 is achieved by using the p-value from Fisher’s exact test as the predictive score. For specific types of functions, AUC of 0.81 is achieved for Mendelian diseases, 0.74 for GWAS phenotypes, 0.85 for pharmacogenetic pathways, 0.76 for GO biological processes, 0.76 for Reactome signaling and metabolic pathways, and 0.72 for GO molecular functions (see **Fig. S6**). The true performance is likely higher, given the function-target function mapping issue.

Not only are target functions of TFs predictive of their known functions, the numbers of target functions (i.e. multi-functionality) are also predictive of the numbers of known functions (Wald t statistic 6.07, p-value $3.1 \times 10^{-9}$; or Wald t statistic 5.07, p-value $6.3 \times 10^{-7}$ after controlling for the number of target genes per TF).

We manually validated the TF-target function relationships for Mendelian diseases, GWAS phenotypes, and pharmacogenetic pathways, for which solid genetic evidence such as direct mutation of the TF in patients are available.

## Mendelian Diseases Targeted by Transcription Factors

We identified the target Mendelian diseases of a TF based on the enrichment of disease causing genes (Hamosh et al. 2005) in the target genes of the TF. In total 181 TF - target Mendelian disease relationships were identified at false discovery rate of 0.05. The majority of the top twenty TF-Mendelian disease associations (from thirteen TFs) are supported by direct genetic evidence such as mutations of the TF in the target Mendelian disease, GWAS associations between the TF and closely related diseases, or phenotypes closely related to the target disease as observed in mouse knockout models of the TF (**Table 1**).

For example, we identified porphyria as a target disease of GATA1 (odds ratio 170, p-value 9.8 x $10^{-9}$), while direct mutation of GATA1 (R216W) has been reported in a congenital erythropoietic porphyria patient (Phillips et al. 2007), and the mutant was suggested to cause the disease by regulating UROS, a common causal gene of congenital erythropoietic porphyria. We identified hereditary mitochondrial metabolism disease as a target disease of NHF4A, while it has been reported that HNF4A is mutated in hereditary noninsulin-dependent diabetes mellitus (NIDDM) (Hani et al. 1998), which is a mitochondrial disease. HNF4A is also mutated in maturity-onset diabetes of the young (MODY1) (Wang et al. 2000), and the MODY1 mutant form of HNF4A is known to cause mitochondrial function defects as well as impairment of nutrient-stimulated insulin release. Transcription factor SOX10 is predicted to be associated with Waardenburg's syndrome (**Table 1**) as well as Hirschsprung's disease and megacolon (not shown). Notably, SOX10 is a known causal gene of Waardenburg's syndrome type 4C and Hirschsprung's disease (Pingault et al. 1998), while megacolon is a symptom of Hirschsprung's disease. Another example is the association of RFX2 with Bardet-Biedl syndrome (BBS), which is a pleiotropic recessive genetic disorder belonging to a group of diseases called ciliopathy (Hildebrandt et al. 2011). There are at least sixteen known BBS causing genes (or nineteen according to OMIM) covering around 80% of the diagnosed cases (Forsythe and Beales 2013). These BBS causing genes encode proteins involved in cilia biogenesis and function. Transcription factor RFX2 is not known to mutate in BBS patients, however, RFX2 together with RFX3 and RFX4, are key regulators of cilia genesis in mouse and other animal models (Bisgrove et al. 2012; Chung et al. 2012).

In addition to above examples for GATA1, HNF4A, SOX10 and RFX2, other target Mendelian diseases among the top twenty are directly supported by human or mouse phenotypes upon direct mutation of the TFs. They are heart septal defect and congenital heart diseases for CTBP2, heart septal defect for SUZ12 (**Fig. 3C**), cancers for ETS1 and TP53, and inherited metabolic disorder for USF1.

## Complex Phenotypes Targeted by Transcription Factors

We identified twenty significant complex phenotypes for seven transcription factors (**Table S3**). Transcription factors NFKB1 and RFX5 (**Fig. 3D**) are each associated with three and six autoimmune disorders, while both TFs are known to be involved in autoimmunity (Baeuerle, Patrick A. 1997; Masternak et al. 2000, 1998). Especially, NFKB1 is recently identified as a causal gene of autosomal dominant variable immunodeficiency-12 (Fliegauf et al. 2015), which shows features of autoimmunity. NFKB1 is also genetically associated with autoimmune disease Ulcerative colitis (Jostins et al. 2012). RFX5 mutations cause Bare lymphocyte syndrome II through impaired MHC II protein expression (DeSandro et al. 1999; Reith and Mach 2001), while MHC II is genetically linked to multiple autoimmune diseases (Fernando et al. 2008). SREBF1 (sterol regulatory element binding transcription factor 1) is identified to target blood LDL (low density lipoprotein) level (p-value $2.2 \times 10^{-6}$), and not surprisingly, SREBF1 is a known regulator of LDL proteins and genes involved in sterol synthesis (Brown and Goldstein 1997; Shimano et al. 1997). Consistently, SREBF1 is associated with the statin pharmacodynamics pathway (PharmGKB) and multiple biological processes and pathways related to lipid metabolism. The SREBF1 LDL association is also directly supported by mouse mutation phenotypes (Hua et al. 1996; Shimano et al. 1997).

Habitual coffee consumption is a significant target phenotype of AHR (Aryl hydrocarbon receptor, **Fig. 3E**), while *AHR* gene locus itself is also strongly associated with coffee consumption and habitual coffee consumption in GWAS studies (Cornelis et al. 2011, 2015; Sulem et al. 2011). HNF1A (**Fig. 3G**) targets three phenotypes, including the serum cell-free DNA level as an indicator of cardiovascular disease risk, serum bilirubin levels as a measure of cholelithiasis risk, and F-cell levels as an indicator of sickle-cell anemia. The first two phenotypes are validated by genetic evidence. HNF1A mutation is associated with elevated risk of cardiovascular diseases in diabetes patients (Steele et al. 2010) while SNPs of HNF1A is found to be associated with cardiovascular disease risks in young and old European Americans (Reiner et al. 2009). The HNF1A-cholelithiasis connection is supported by HNF1A knockout mouse (Pontoglio et al. 1996), which showed elevated blood bilirubin levels and jaundice. Consistently, HNF1A also targets Reactome pathway *synthesis of bile acids and bile salts* (**Fig. 3G**).

## Pharmacogenetic Pathways Targeted by Transcription Factors

We identified 99 TF-target pharmacogenomic pathway relationships, covering 47 unique TFs and 45 unique pharmacogenetic pathways in PharmGKB. There is no preference towards pharmacokinetics (PK) or pharmacodynamics (PD) pathways, with 20 of 40 PK pathways and 26 of 50 PD pathways identified. However, different TFs are responsible for the target PK and PD pathways. 18 of the 26 target PK pathways are the targets of just 4 TFs (see **Table S4**), i.e. HNF1A, AHR, NR1I3, and FOXA2. Among them, nuclear receptor genes HNF1A, AHR, and NR1I3 are well known to regulate xenobiotic-metabolizing enzymes (Pontoglio et al. 1996; Sogawa and Fujii-Kuriyama 1997; Lamba et al. 2005; Ma 2008). Unique target PK pathways are found for each of the four TFs, suggesting their complementary roles in regulating drug metabolism. In addition to these 4 TFs, SP1 and TP53 are each associated with 3 PK pathways for cancer drugs. SP1 and TP53 are also associated with other cancer PD pathways, and their associations with cancer are strongly supported by the literature (Li and Davie 2010; Hollstein et al. 1991).

We manually examined the full list of identified target PD pathways and confirmed majority of the associations (see **Table S4)**. A PD pathway describes the disease pathway that is perturbed by a drug. A target PD pathway is considered confirmed if the TF is a member of the PD pathway or closely related pathways, or if the TF is known to be genetically linked to the disease or closely related phenotypes. For example, ELK1 is identified as a regulator of the EGFR Inhibitor Pathway, while the TF itself is a member of the PD pathway. HNF1A is identified as a regulator of the PD pathways for cancer, high cholesterol, and diabetes, while mutations of HNF1A is known to cause hereditary cancers and diabetes, and variants of HNF1A are strongly associated with cholesterol level in GWAS (Teslovich et al. 2010). E2F1 and E2F4 are identified for multiple antimetabolite PD pathways. Antimetabolites are a class of drugs for inducing medical abortions and treating cancers and autoimmune diseases through halting the cell cycles, while E2F1 and E2F4 are well-known regulators of cell cycles (Ren et al. 2002; Gaubatz et al. 2000).

## Target Gene Sharing among Transcription Factors

While the target genes of a transcription factor define its biological functions, the target gene sharing between two TFs also reflects the functional relatedness between TFs. We studied the relationship between the target gene overlaps and target function sharing between pairs of TFs.

As expected, the target gene sharing, measured by Pearson's phi coefficient $\phi_{TG}$, is highly associated with the target function sharing $\phi_{Target\ Fun}$ (Wald T-statistic 126.95, or 109.75 when controlling for the number of target genes, both p-values < $2.2 \times 10^{-16}$). Among 73,536 possible TF pairs (**Fig. S7)**, 12,434 (16.9%) show significant target gene sharing at false discovery rate of 0.01 based on Fisher's exact test. We refer to these similar TFs as *TF neighbors*. Relatedly, there are 11,205 pairs of TFs with one or more shared target functions, including 5,866 pairs that also show significant target gene sharing (odds ratio 9.3).

Despite the overall consistency between target function overlap and target gene sharing, many exceptions occur. Significant target gene sharing was observed for 6568 pairs of TFs that did not share any target functions, including 428 pairs that surprisingly showed negative correlations between their target function association profiles[1]. This could be caused by unknown or poorly understood functions common to these TF neighbors, and it suggests that the target gene-based TF neighbors may provide functional information missed by the target functions, therefore, the TF neighbors serve as an additional layer in the TF function annotations. On the other hand, significant target function sharing (at FDR < 0.05) is observed for 329 pairs of TFs that have lower-than-expected target gene overlaps. The top five TF pairs by target function sharing are MXI1 and RFX1, TRIM28 and VDR, LMO2 and ZNF263, ARNTL and BHLHE40, ETV5 and MXI1. Among these, two pairs, TRIM28 and VDR, ARNTL and BHLHE40, do not share any target genes. However, TRIM28 and VDR (Vitamin D Receptor) share 12 target functions, e.g. *Reactome Telomere Maintenance*, out of 15 and 14 target functions for the two TFs respectively; while ARNTL and BHLHE40 share 2 target functions, *Reactome Bmal1 Clock Npas2 Activates Circadian Expression* and *Reactome Circadian Clock*, out of 4 and 2 target functions for the two TFs respectively. Mouse gene knock out confirmed the abnormal circadian rhythm as a phenotype for both ARNTL (Bunger et al. 2000; Storch et al. 2007) and BHLHE40 (Rossner et al. 2008), and the two proteins may be interaction partners (Honma et al. 2002).

[1] The target function association profile of a TF is comprised of the Pearson's phi coefficients between the TF and all 3715 functional concepts. A lack of positive correlation between two profiles indicates that the two TFs are likely functionally unrelated based on the known functional concepts.

## Transcription Factors Link Apparently Unrelated Functions: Coffee and Warfarin

Parallel to the TF target gene sharing and TF target function sharing, we observed extensive member gene overlaps and regulator sharing between pairs of functional concepts (see Supplemental Material Section 1.3). Majority of TFs (64%) have two or more target functions. We observed that apparently unrelated gene functions are frequently linked by transcription factors. For example, AHR is found to be associated with coffee consumption, the PK pathways for drugs Amodiaquine, Warfarin, Erlotinib, and Phenytoin, as well as the estrogen metabolism pathways (**Fig. 3E**). Based on these observations, we hypothesize that coffee consumption would interfere with the metabolism of these drugs and estrogen, either through modifying the activities of AHR target enzymes or impacting the expression of the enzyme genes through feedback regulation of AHR activity. The interactions of coffee drinking with both Warfarin (Zambon et al. 2011) and Phenytoin (Wietholtz et al. 1989) have been reported. On the other hand, coffee consumption is actually associated with decreased venous thromboembolism (Enga et al. 2011), which warfarin can effectively treat. The coffee-estrogen link is even more intriguing. High coffee intake is found in multiple studies to be significantly associated with decreased risk of estrogen-receptor negative breast cancer (Li et al. 2011; Lowcock et al. 2013) and breast cancer risk in BRCA mutant carriers (Nkondjock et al. 2006). In addition, high coffee intake impacts the risk of Parkinson's disease in female in an estrogen dependent manner (Ascherio et al. 2003, 2004), possibly through modifying blood estrogen levels (Nagata et al. 1998).

Obviously, when two functional concepts are statistically associated, i.e. when they share significant number of member genes, they will likely be linked to the same regulators (**Fig. 4A** and **Fig. S9A)**. The inverse is however not true. Two functions can be linked by transcription factors even when they do not share significant portion of member genes (**Fig. 4B** and **Fig. S9B**). In fact, of the 954 function pairs that share identical sets of regulators, 356 (37%) pairs have less gene overlap than expected by chance (**Table S7**), i.e. with odds ratio < 1. Most of such function pairs do not share any member genes. For example, hereditary “lipid storage diseases” do not share any genes with Reactome pathways “iron uptake and transport” and “insulin receptor recycling”, but the 3 functions are found to share regulators ATF3, NFE2, USF1, and USF2, while “iron uptake and transport” is also a target function of ARNT (**Fig. 4C**). Other examples include *ventricular septal defect* and developmental *pattern specification process* which are both targeted by SUZ12 and CTBP2, *PECAM1 Interactions* and disease *agammaglobulinemia* targeted by EBF1, *intestinal disease* and *Human immunodeficiency virus infectious disease* both targeted by NFKB1, *prostate cancer* and *intestinal cancer* both targeted by TP53, *Metalloendopeptidase Activity* and *cognitive disorder* both targeted by ETV4, among many others.

## Measuring the Functional Pleiotropy of Transcription Factors

A transcription factor is functional pleiotropic if it targets multiple unrelated functions. The above analyses suggest extensive functional pleiotropy for transcription factors. Here, we quantify the functional pleiotropy of TFs in order to further study its causes from the perspective of transcriptional regulation. The number of target functions $n_{Target\ Fun}$ can be a measure of TF functional pleiotropy, with the caveat that it double counts closely related or redundant functional concepts. We hence define *function diversity* $\pi_{Target\ Fun}$ as the “effective” number of target functions by weighting each function by its *uniqueness*, which is the inverse of the accumulative similarity between the function and other functional concepts. Similarly, we define *regulator diversity* $\pi_{Reg}$ of a gene as the effective number of regulators. The regulator diversity corrects for related or cooperative transcription factors that are counted independently in the number of regulators $n_{Reg}$ targeting a gene (see Methods and Supplemental Material Section 1.4 for details).

The two most functional pleiotropic TFs are BRCA1 and ZNF143. They are annotated with 272 (45.5 effectively) and 242 (35.4 effectively) target functions, and 101 (14.5) and 22 (2.8) known functions, while regulated by 33 (15.9) and 53 (22.1) upstream TFs respectively. The two TFs of the highest upstream regulatory diversity are MYC and TP53. They have 50 (25.1) and 49 (23.0) regulators, and are annotated by 159 (24.2) and 175 (26.8) target functions, and 68 (12.4) and 166 (25.4) known functions respectively. By contrast, TFs with smaller number of target functions also tend to have fewer upstream regulators. For example, HNF1A has 9 (3.8) regulators, 30 (6.3) target functions, and 58 (11.2) known functions (**Fig. 5A**); NFKB1 has 26 (11.7) regulators, 143 (23.7) target functions, and 34 (6.2) known functions (**Fig. S11A**); SUZ12 has 11 (4.9) regulators, 48 (8.3) target functions, and 4 (1.0) known functions. **Fig. 6** provides the target gene-based function annotations for TFs NFKB1 and SUZ12, including the target functions, TF neighbors, as well as the function diversity measures.

## Upstream Regulation Enables Functional Pleiotropy of Transcription Factors

Over the set of 384 TFs in the TFTG compendium, we observed a global positive association between the target function diversity of TFs with the regulator diversity (Wald test p-value $3.3 \times 10^{-10}$ between diversity measures $\pi_{Target\ Fun}$ and $\pi_{Reg}$; or p-value $1.6 \times 10^{-9}$ between raw counts $n_{Target\ Fun}$ and $n_{Reg}$), i.e. TFs with more effectively-unrelated upstream regulators also tend to have more effectively-unrelated target functions. This suggests diverse modes of upstream regulation as a mechanism for TFs to realize functional pleiotropy. To eliminate technical biases due to ChIP-seq experiment quality or uneven research attention for different TFs, we controlled for $n_{TG}$, the number of target genes per TF, as a confounding factor through linear model. Regulator diversity however remained a significant predictor of TF's function diversity (p-value $5.3 \times 10^{-6}$ , Wald test). Further, we examined the known functions of TFs, which, unlike the target functions, are independent of the TFTG data compendium. A significant association remained between the known function diversity and the regulator diversity of TFs (p-value $6.3 \times 10^{-5}$ between diversity measures $\pi_{Known\ Fun}$ and $\pi_{Reg}$, or p-value 0.00022 between raw counts $n_{Known\ Fun}$ and $n_{Reg}$). This was true regardless of the number of target genes for the TFs. In fact, slightly stronger correlation was observed when TFs with less than 100 target genes were removed (**Fig. S12**). Finally, to completely eliminate the impact of human research biases toward popular TFs, which could result in higher number of literature-reported target genes as well as literature-reported upstream regulators for the popular TFs, we repeated the above all experiments after removing all low-throughput (literature derived) data in the TFTG compendium. We observed that regulator diversity and function diversity remain significantly associated (see Supplemental Material Section 1.5). As a control, we evaluated the association between TF's Functional Pleiotropy and its hierarchical location within the gene regulatory network, measure by PageRank (Page et al. 1999). Neither the PageRank-function diversity nor the PageRank-target function diversity associations are significant after controlling for the number of target genes of TFs (see Supplemental Material Section 1.6).

In addition, we observed that the positive association was universal for all six types of function annotations. The trends are stronger for biological processes and molecular pathways, and weaker for GWAS and disease phenotypes (**Fig. 5B**). The association between function and regulator diversities extends to non-TF genes as well, with p-value $7.9 \times 10^{-5}$ between diversity measures $\pi_{Fun}$ and $\pi_{Reg}$ and p-value $3.0 \times 10^{-18}$ between raw counts $n_{Fun}$ and $n_{Reg}$ for 11345 genes that have both regulator and function annotations (see Supplemental Material Section 1.7).

If regulator diversity is indeed a cause of TF function diversity, it is likely through driving the expression of the TF in diverse conditions. To evaluate this mechanism, we examined the expression of transcription factors in a collection of 327 human tissue types and cell lines (McCall et al. 2011). As expected, expression diversity of TFs is significantly associated with the regulator diversity (Spearman rank correlation 0.22, p-value $2.7 \times 10^{-6}$, or Spearman rank correlation 0.26, p-value $3.6 \times 10^{-7}$ for the raw counts). On the other hand, there is a significant association between expression diversity of TFs and the target-function diversity (Spearman's rank correlation 0.10, p-value 0.048), and the function diversity (Spearman's rank correlation 0.26, p-value $2.2\times 10^{-7}$). Similarly, we observed strong associations between expression diversity of general genes and the function and regulator diversities of genes (see Supplemental Material Section 1.7). These support transcriptional regulation diversity as a mechanism for functional pleiotropy of transcription factors and other genes.

# Discussions

A major challenge in data-driven TF function annotation is to minimize the impacts from false bindings and to reliably extract gene function signals. We combined multiple statistical strategies to achieve this. First, we target genes from ChIP-seq experiments are extracted with a stringent false discovery rate (FDR), which is calculated using a statistical framework modified from TIP (Cheng et al. 2011a) that combines binding locations and intensity information to differentiate true TF-DNA binding events from false signals. Second, we define the target functions of TFs as the consensus functions among the putative target genes. The statistical enrichment analysis hence further filters noises from the remaining false target genes. Third, we choose conservative gene universes specific to the types of functions, so as to minimize spurious associations. Finally, we apply the Benjamini–Hochberg multi-test correction procedure and require a FDR of 5% for all associations we report. With these, around 10K significant TF-target function associations are obtained. Meanwhile, the total number of true TF-target function associations is estimated to be over 80K, indicating the presence of rich functional signals in the TFTG data (**Fig.3**). We believe there is room for further improvement to retrieve higher number of TF-target function annotations at controlled false discovery rate.

We globally validate the TF-target function associations by comparing them with known TF-function relationships, and show that the target functions cover both known and novel TF-function relationships. Despite the fact that TF-target function and TF-function relationships do not always have direct correspondence, we observe a good prediction performance with AUC 0.80 with 6 types of gene functions combined. In addition, we manually validate the top target diseases, phenotypes, and pharmacogenetic pathways based on the literature, and find that majority of them are supported by direct genetic evidence, such as direct mutations or GWAS implicated associations of a TF in patients with the target disease, or phenotypes of mouse knock-out models of the TF (**Table 1**, **S3** and **S4**), even when they are not annotated as known function of the TFs. Given that our knowledge is incomplete for even the most well studied TFs, we believe the non-validated TF-target functions represent opportunities for future experimental studies of the TFs.

The foundation of this study is the hypothesis that genes regulated by a same TF are functionally related. We believe this extends to the functional concept level, i.e. multiple concepts targeted by the same TF(s) are also functionally related at some higher level. Based on co-regulation, we predict the interaction between coffee consumption and the metabolism of multiple drugs including warfarin as well as the interaction between coffee consumption and estrogen metabolism, both of which are validated by multiple published experimental studies (Wietholtz et al. 1989; Zambon et al. 2011; Enga et al. 2011; Li et al. 2011; Lowcock et al. 2013; Nkondjock et al. 2006; Ascherio et al. 2003, 2004). Further, we show that TFs link hundreds of functional concept pairs that do not share any member genes. This highlights the potential usage of the TF-target function network to study the high-level organization principles among biological functions that is unattainable by solely studying the member genes of functions, e.g. through a member gene-based function-function association network.

Based on the TF-target function network, we examine the functional pleiotropy of TFs. We discover that a TF with more target functions (or known functions) are themselves regulated by significantly more TFs, and both function and regulator diversities are associated with the expression diversity of the TF in cell lines and tissues. These suggest that regulator diversity may be a cause of function diversity of TFs, and it works by driving the expression diversity of genes.

Gene regulation is well known to be cell type specific, and co-expression of TFs is required for the co-regulation of TFs on the shared target genes (Ravasi et al. 2010). However current high-throughput studies for in-vivo TF-DNA binding, including the ENCODE project (Wang et al. 2013; Gerstein et al. 2012), are generally limited to a small number of tissue/cell types. Comprehensive ChIP-seq analysis on a large number of cell types remains unrealistic due to cost and resource requirements. We hence compile the transcription factor - target gene relationships in a cell type and development stage agnostic manner. Contingent on data availability, this work can be easily extended to perform cell type specific TF function annotation. Despite this compromise, the resulting TFTG data turns out to partially capture the cell type specificity of TFs, as we observe TFs sharing similar tissue expression patterns also share greater amount of target genes (Wald T-test, p-value $1.9 \times 10^{-78}$).

In an effort to manually annotate TF functions, (Yusuf et al. 2012) teamed up over 100 experts to curate and integrate published knowledge and provide mini-reviews on transcription factors. We believe automated yet accurate function annotation and manual curation are complementary and will together greatly facilitate our understanding of the biological functions of human transcription factors.

Despite large consortium efforts such as ENCODE (Bailey et al. 2013; Gerstein et al. 2012), existing data for transcription factor-target gene relationships remains scarce. Our TFTG compendium covers 384 unique transcription factors. This is the largest collection to our knowledge, as compared to 237 TFs in a recently published study (Griffon et al. 2015), yet this only covers 20-25% of the putative 1500-2000 transcription factors in human (Vaquerizas et al. 2009; Kummerfeld and Teichmann 2006). Relatedly, we notice that the TFTG compendium is biased toward the well-known transcription factors, likely due to preferential attachment of research efforts to popular TFs. For the same reason, some TFs enjoy higher target gene coverage than the other TFs. These biases currently limit the power of target gene based TF function annotation. However, with the maturity of ChIP-seq and related high-throughput assays for in-vivo protein-DNA binding and the availability of the technologies to more labs, we expect a stead accumulation of TFTG data with improved accuracy and completeness yet reduced biases. Such data will ultimately enable the annotation of all transcription factors in the human genome, and serve as the foundation for hypothesis generation and further experimental studies of the roles of TFs in normal biological processes and diseases.

# Methods

## Transcription Factor Target Gene Data Compendium

We compiled transcription factor-target gene (TFTG) relationships from multiple sources. ChIP-seq experiments from both large-scale (Bernstein et al. 2012; Gerstein et al. 2012; Yan et al. 2013) and small-scale studies were included. Meta-data of 413 ChIP-seq experiments for 235 unique transcription factors were curated manually (by October 2013) from GEO (Barrett et al. 2013), in addition to around 450 ChIP-seq experiments for 115 unique transcription factors from the ENCyclopedia of DNA elements (ENCODE) (Bernstein et al. 2012; Gerstein et al. 2012). Manually curated low-throughput target gene annotations were compiled from multiple databases, including BIND, HTRI, PAZAR, and TRED (Bader 2003; Bovolenta et al. 2012; Portales-Casamar et al. 2009; Jiang et al. 2007). Only TFTG relationships with direct literature evidence from low throughput experiments (Yang 1998; Geertz and Maerkl 2010), e.g. as electrophoretic mobility shift assays, were included. We did not differentiate sequence specific DNA-binding transcription factors from other DNA binding transcriptional regulators. Some cofactors that do not directly bind DNA are also included when there are ChIP-seq data available. Despite this, we refer to all these transcriptional regulators as transcription factors (TFs) in this study. The binding signals from target genes were differentiated from that from non-target genes using a modified version of the TIP algorithm (Cheng et al. 2011a), which combines the binding location and intensity information for scoring TF target genes (See Supplemental Methods section 2.1).

## Gene Function Annotation Data

Six types of gene annotations were used in this analysis to annotate transcription factors. Gene Ontology (GO) (Ashburner et al. 2000) for biological processes (BP) and molecular functions (MF), together with the Reactome pathways (Joshi-Tope et al. 2005) were retrieved from the MSigDB v4.0 (Liberzon et al. 2011). The Pharmacogenomics pathways for pharmacodynamics (PD) and pharmacokinetics (PK) were retrieved on Jan. 20, 2013 from the Pharmacogenomics Knowledgebase (PharmGKB) (Hewett 2002). Gene disease association data from genome wide association studies (GWAS) were obtained on May 4, 2014 from dbGAP (Mailman et al. 2007) and NHGRI (Welter et al. 2014) catalogs with p-value cutoffs at 1E-3 (loose set) or 1E-5 (stringent set), and the closest gene, or two genes if the SNP is intergenic, to each SNP is retained. When not specified, the loose set is used. Note here a large p-value cutoff is used to capture majority of the true disease related genes rather than to select for confident ones, as our goal here is to associate complex phenotypes and diseases rather than individual genes with transcription factors. The gene - Mendelian disease annotations were obtained on July 5, 2014 from the Online Mendelian Inheritance in Man (OMIM) (Hamosh et al. 2005), and the disease genes were further grouped in a hierarchical manner to disease classes based on the disease ontology (Schriml et al. 2012). For all data, only genes uniquely mapped to the Entrez Gene database (Maglott et al. 2005) are retained.

## Defining Coding Genes and Literature Rich Genes

Coding genes are defined as all Entrez genes that have associated protein products in Ensembl Protein or UniProt databases. Literature Rich genes are defined as coding genes annotated in any of the following 7 data sources: GO Biological Processes, GO Molecular Functions, Reactome, PharmGKB, Kyoto Encyclopedia of Genes and Genomes pathways(Kanehisa 2000), Biocart(Nishimura 2001), and OMIM(Hamosh et al. 2005). There are 19847 Coding and 10931 Literature Rich genes in total. Interestingly, 333 of the Literature Rich genes are not Coding genes, but pseudogenes, discontinued gene records, or gene loci without defined genes. These were removed, leaving 10561 Literature Rich genes in total.

## Measuring the Associations between Binary Variables

Fisher's exact test (Mehta 1986) is used for testing the associations between transcription factors and biological functions, by detecting significant enrichment of genes that are target genes of a TF and are also annotated with a given function. G-test is used as a fast approximation to Fisher's exact test in preliminary analyses and to demonstrate the presence of functional signals in the TF target gene data (Figure 3). To perform multi-test correction, we calculated the Benjamini-Hochberg false discovery rate (Benjamini and Hochberg 1995) on the p-values for each type of annotation separately.

Since Fisher's exact test does not have a test statistic that we can use to measure the similarities between two binary variables. We use Pearson's phi coefficient ($\phi$, PPC) to measure association strength,

$$\phi=\frac{n_{11}n_{00}-n_{10}n_{01}}{\sqrt{(n_{10}+n_{11})(n_{00}+n_{10})(n_{01}+n_{11})(n_{00}+n_{01})}},$$

where $n_{ij}$ are the observed number of $ij$ value pairs for the two random variables. The strengths of TF-Function association, TF-TF target gene sharing, TF-TF target function sharing, TF-TF known function sharing, and Function-Function member gene sharing are denoted as $\phi_{TF_Fun}$, $\phi_{TG}$, $\phi_{TargetFun}$, $\phi_{Fun}$, and $\phi_{Fun_Fun}$ respectively. PPC is sample size independent, and serves as a good measure of the magnitude of associations. The sign of PPC indicates the directionality of an association.

## Functional and Regulator Diversities of Transcription Factors

We measure the effective number of transcription factors (i.e. the regulatory diversity) of a function or gene and the effective number of target functions (i.e. the function diversity) of a TF by down weighting the TFs (or functions) that are correlated with other TFs (or functions). Given Pearson's phi coefficient $\phi_{tt'}$ between TFs $t$ and $t'$, the uniqueness of TF $t$ is defined as $u_t = 1/\sum_{t' \in TFs} \phi_{tt'}^2$. Note that $u_t$ is always within 0 to 1, since the association between a TF with itself is always 1, i.e. $\phi_{tt}^2 = 1$. The regulator diversity $\pi_{Reg.g}$ of a function or gene (including TF) $g$ is then defined as the weighted counts of the transcription factors targeting the function or gene, $\pi_{Reg.g} = \sum_{t \in TFs\ regulating\ g} u_t$. The regulator diversity measures the effective (non-redundant) number of regulators for a gene (or a TF). Similarly we can define the uniqueness of each function annotation term, phenotype, or disease, and then define the target function diversity $\pi_{Target\ Fun}$ (i.e. effective number of target functions) of a transcription factor or the function diversity $\pi_{Fun}$ (i.e. effective number of known functions) of a gene.

## Data Access

The Cytoscape file for the full TF-function network is available at simtk.org under identifier TFAnno and the full target function annotations for TFs are available as supplemental Table S7.

## Acknowledgement

We thank Fuxiao Xin for her comments on the manuscripts. YFL would like to acknowledge the support of TRAM pilot grant for part of this work. RBA would like to acknowledge funding NIH GM102365, GM61374 and HL117798.

## Figure Legends

Figure 1:

An outline of the workflow for regulatory network based annotation of transcription factor functions.

Figure 2:

Presence of gene function signals in the TFTG data. The scatter plot shows the p-values of function-TF associations obtained using real TFTG compendium (x-axis) and a fake TFTG compendium (y-axis). Each dot corresponds to a pair of p-values for a TF-Function pair. The inlet shows the number of significant TF-Target Function relationships at varying p-value cutoffs for the real TFTG data (y-axis) against the number for the fake TFTG data (x-axis). P-values are obtained by G-tests.

Figure 3:

(A) Global view of the transcription factors and their target functions. 311 TFs and 1420 annotations with one or more significant associations at FDR 0.1 levels are retained. Red indicates positive associations, green indicates negative associations, white indicates FDR > 0.1. Intensity of the colors corresponds to the significance levels: FDR 0.1, 0.05, and 0.01.The TF and target function clustering showed on the left and top was performed based on the TF-target function association phi coefficient matrix. We used the literature rich gene universe for the association analysis except for the TF-GWAS phenotype association, for which the coding gene universe is used.

(B) The network visualization (Smoot et al. 2011) of TF-target function and TF-known function relationships. Edges are colored red or green the same way as in (A). A solid edge links a TF with a significant target function that is not a known function. A dashed edge links a TF with a known function. A dashed edge with color links a TF with a known function that is also a significant target function, while a grey dashed edge links a TF with a known function that is not a significant target function. Node colors and shapes correspond to function types: purple circles, TFs; grey rectangles, Reactome pathways; blue triangles, GO molecular functions; white diamonds, GO biological processes; red rhomboids, PharmGKB PK and PD pathways; yellow hexagon, Mendelian diseases; green octagons, GWAS phenotypes.

Figure 4:

Transcription factor sharing among apparently unrelated functional concepts. (A) Two functional concepts with high member gene overlaps always have similar regulators, but (B) two functional concepts with nearly identical regulators do not always have high member gene sharing. (C) A Venn diagram for 3 functional concepts for which shared transcription factors are identified for functions without gene overlaps. The arrows connect the significant regulators for the functions. Note that “Iron Uptake and Transport” and “Insulin Receptor Recycling” do share member genes significantly, but neither of them shares member genes with “Lipid Storage Disease”.

Figure 5:

The relationship between functional diversity and regulator diversity of transcription factors.

(A) The target functions of transcription factor HNF1A form 3 major clusters based on similarities (member gene sharing) among the functions, while the upstream regulators of HNF1A form clusters based on the functional similarities (target gene overlaps) among these regulators. The regulator and functional diversities of a gene measures the effective number of regulators and effective number of functions for a gene. The coloring schema is same as in Figure 5 and the clustering of TFs and functions are based on the TF's target gene overlaps and function's member gene overlaps.

(B) Significant associations exist between the regulator diversity and target function diversity of transcription factors for six types of function annotations separately.

Figure 6:

Complete target gene-based annotations for two example transcription factors (A) NFKB1 and (B) SUZ12. Three types of information are provided: 1) the top TF neighbors obtained by TF distance (1- target-gene overlap measured by Pearson's phi coefficient) < 0.8; 2) the target functions in six categories, 3) the functional diversities in six categories and total diversity. See Figure S9 for a visualization of the regulator and target function networks surrounding NFKB1.

# Figures

Figure 1:

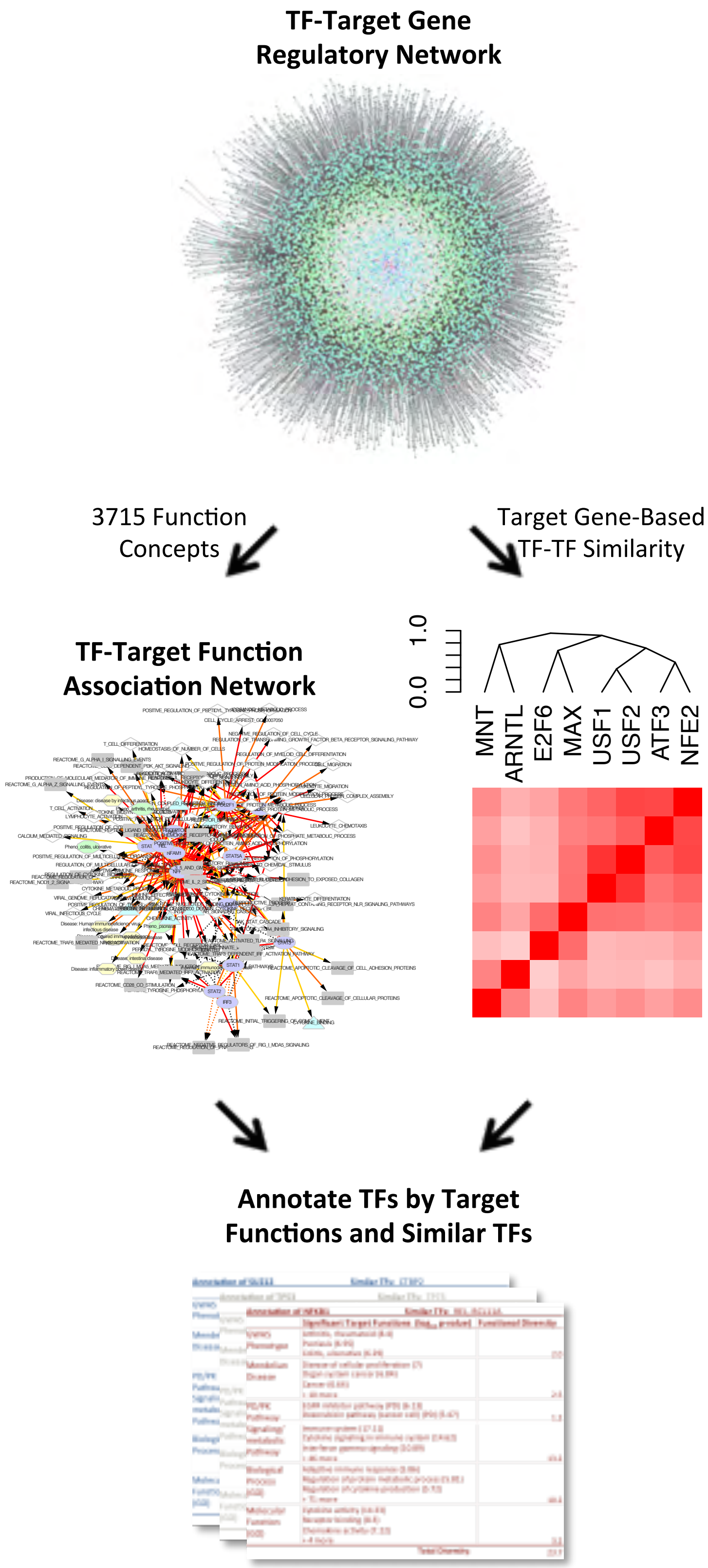

Figure 2:

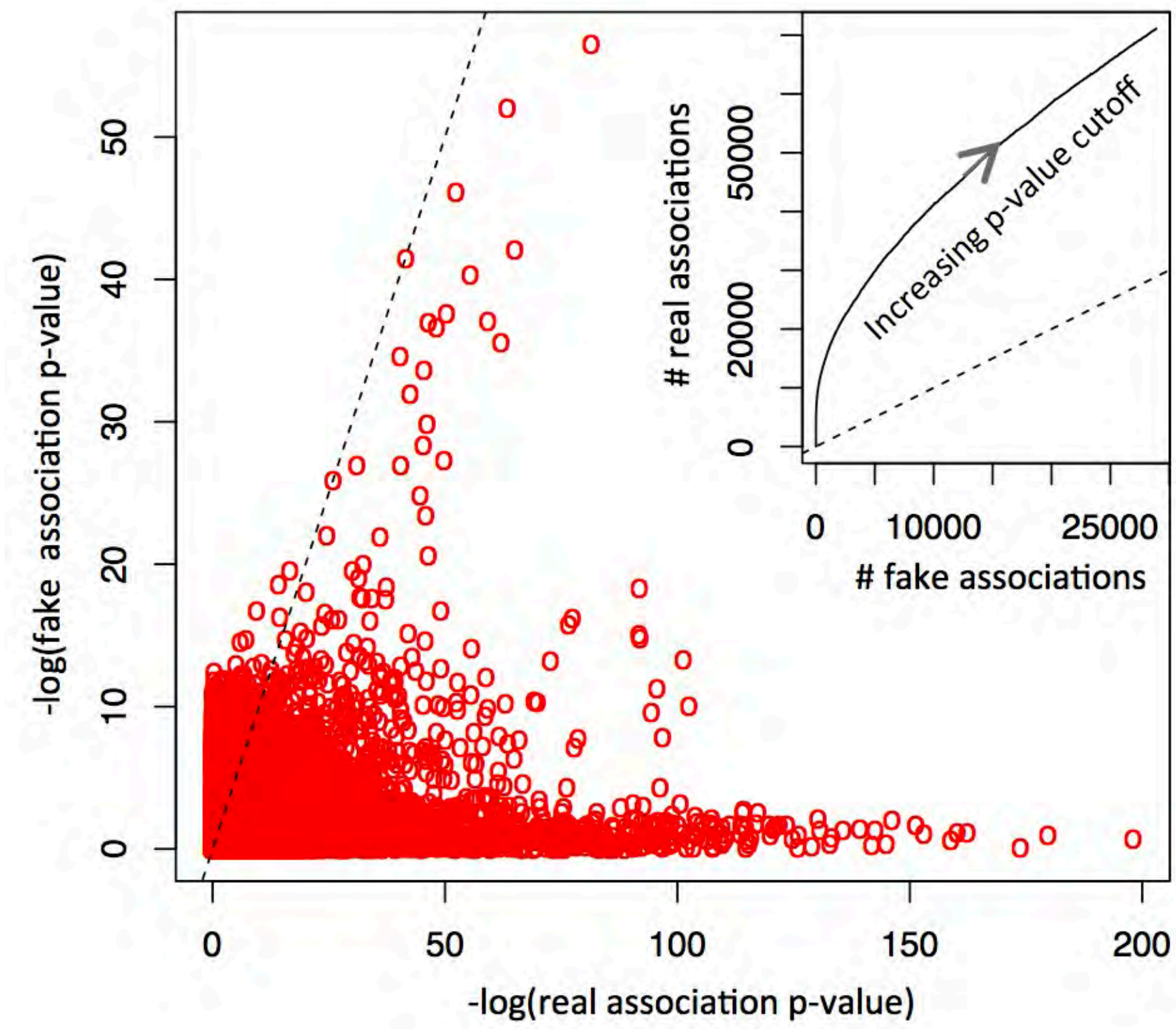

Figure 3:

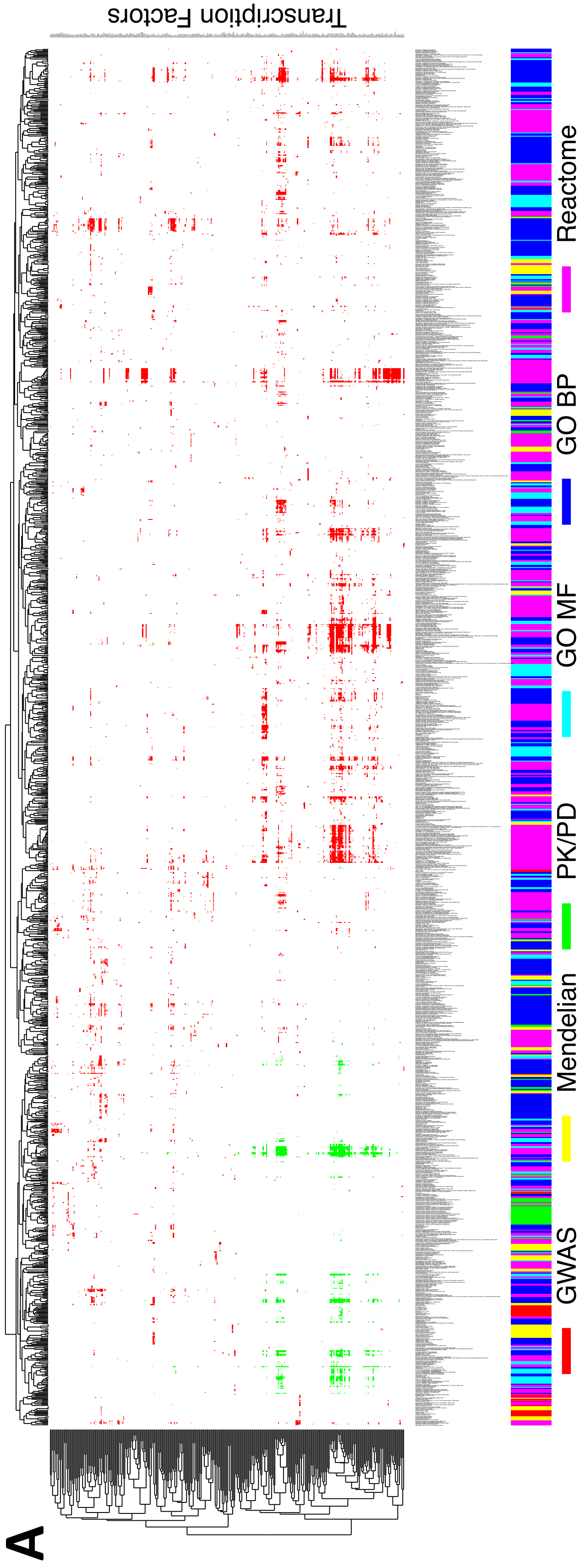

# B Transcription Factor-Function Network

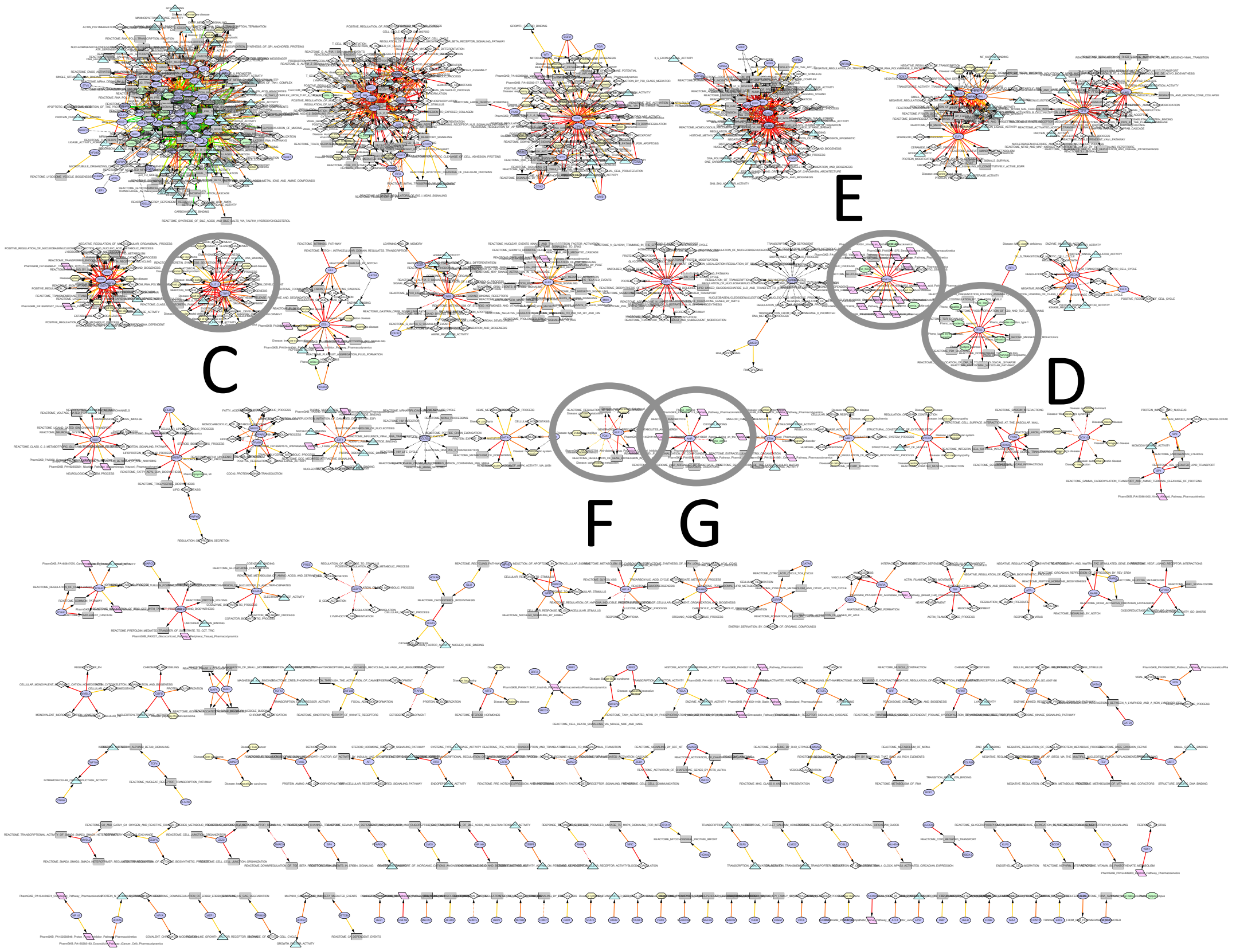

**C**

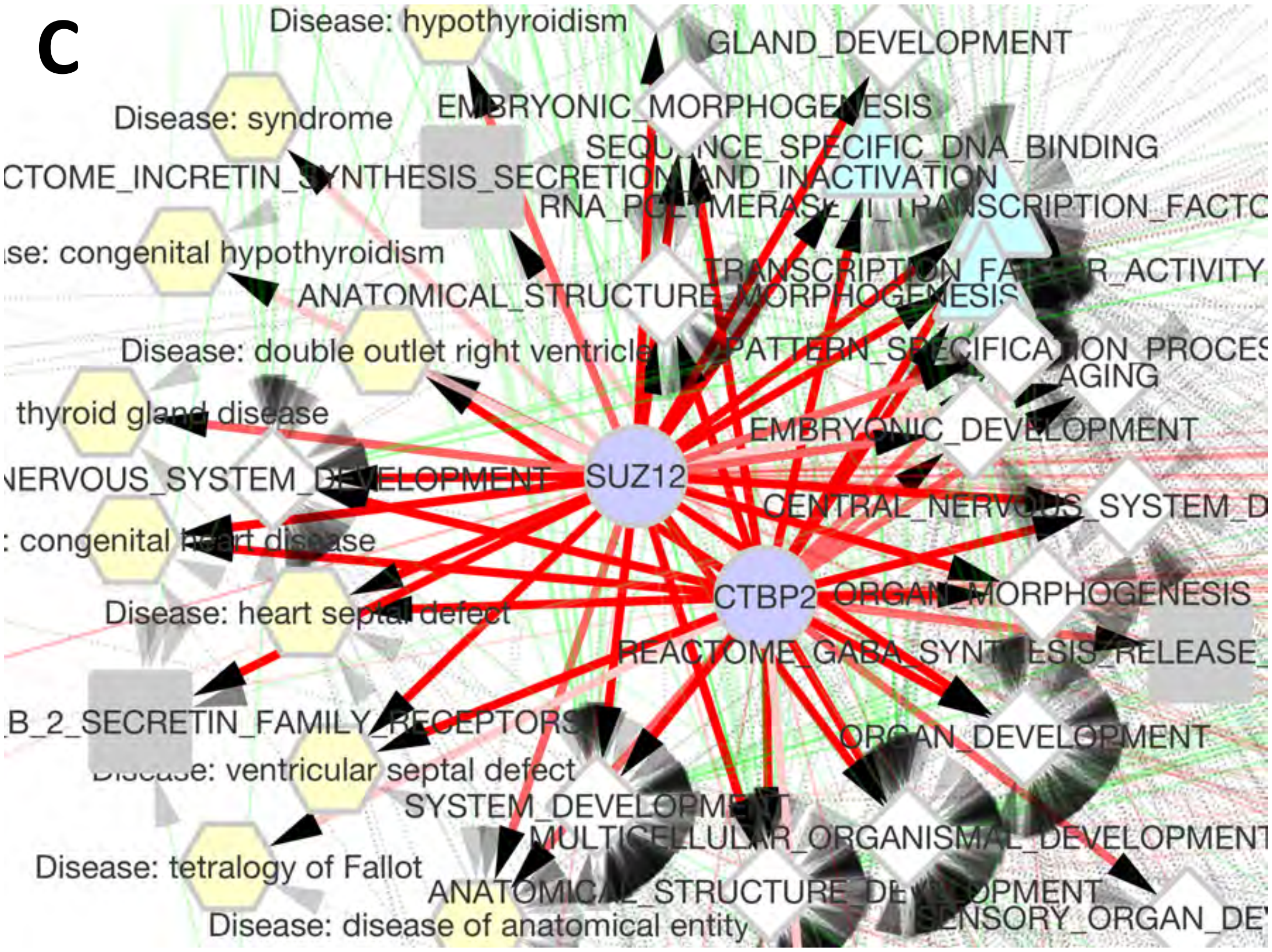

**D**

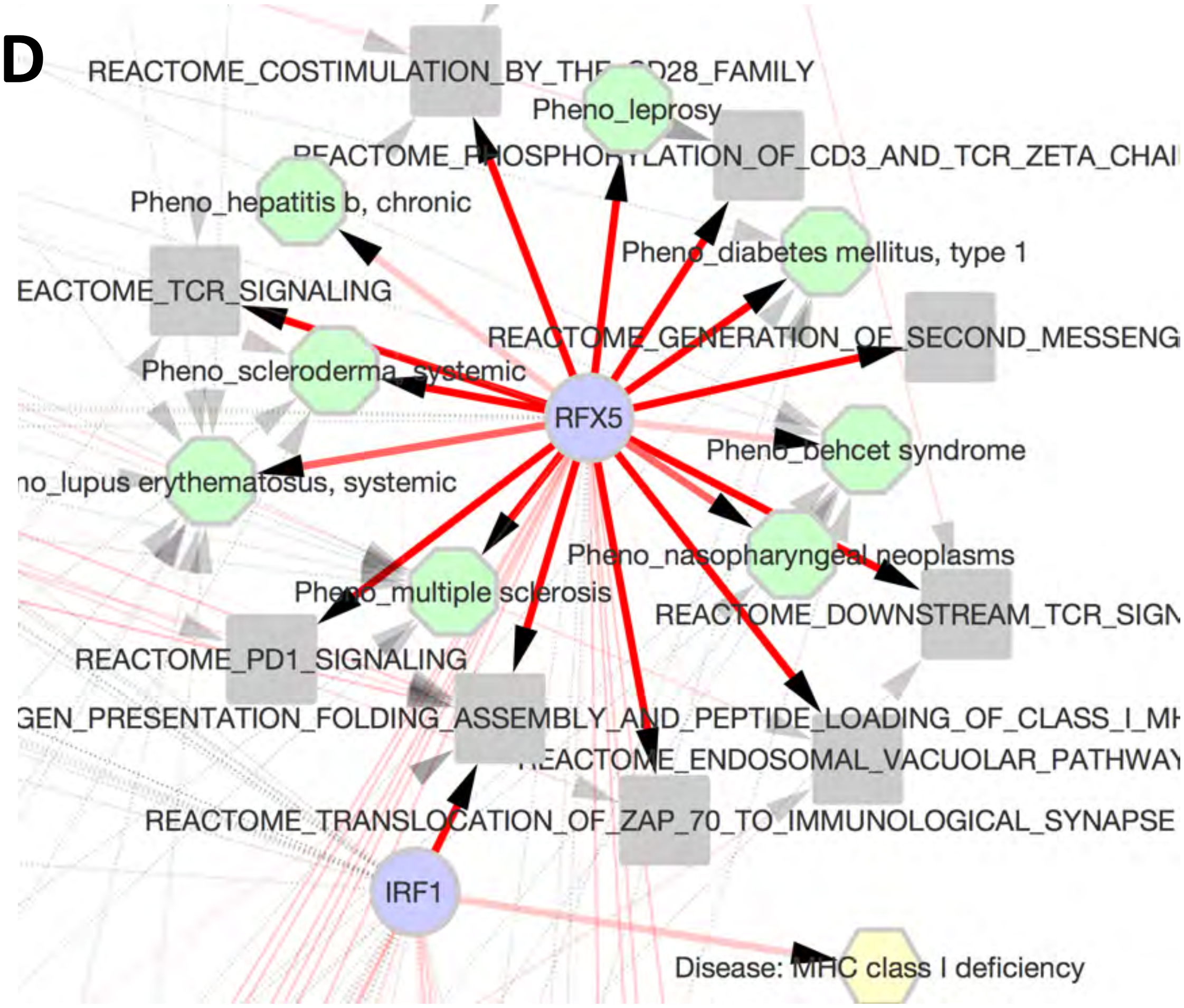
REACTOME_COSTIMULATION_BY_THE_CD28_FAMILY
Pheno_leprosy
REACTOME_PHOSPHORYLATION_OF_CD3_AND_TCR_ZETA_CHAI
Pheno_hepatitis b, chronic
Pheno_diabetes mellitus, type 1
EACTOME_TCR_SIGNALING
REACTOME_GENERATION_OF_SECOND_MESSENG
Pheno_scleroderma, systemic
RFX5
Pheno_behcet syndrome
no_lupus erythematosus, systemic
Pheno_nasopharyngeal neoplasms
Pheno_multiple sclerosis
REACTOME_DOWNSTREAM_TCR_SIGN
REACTOME_PD1_SIGNALING
GEN_PRESENTATION_FOLDING_ASSEMBLY_AND_PEPTIDE_LOADING_OF_CLASS_I_MH
REACTOME_ENDOSOMAL_VACUOLAR_PATHWAY
REACTOME_TRANSLOCATION_OF_ZAP_70_TO_IMMUNOLOGICAL_SYNAPSE
IRF1
Disease: MHC class I deficiency

E

F

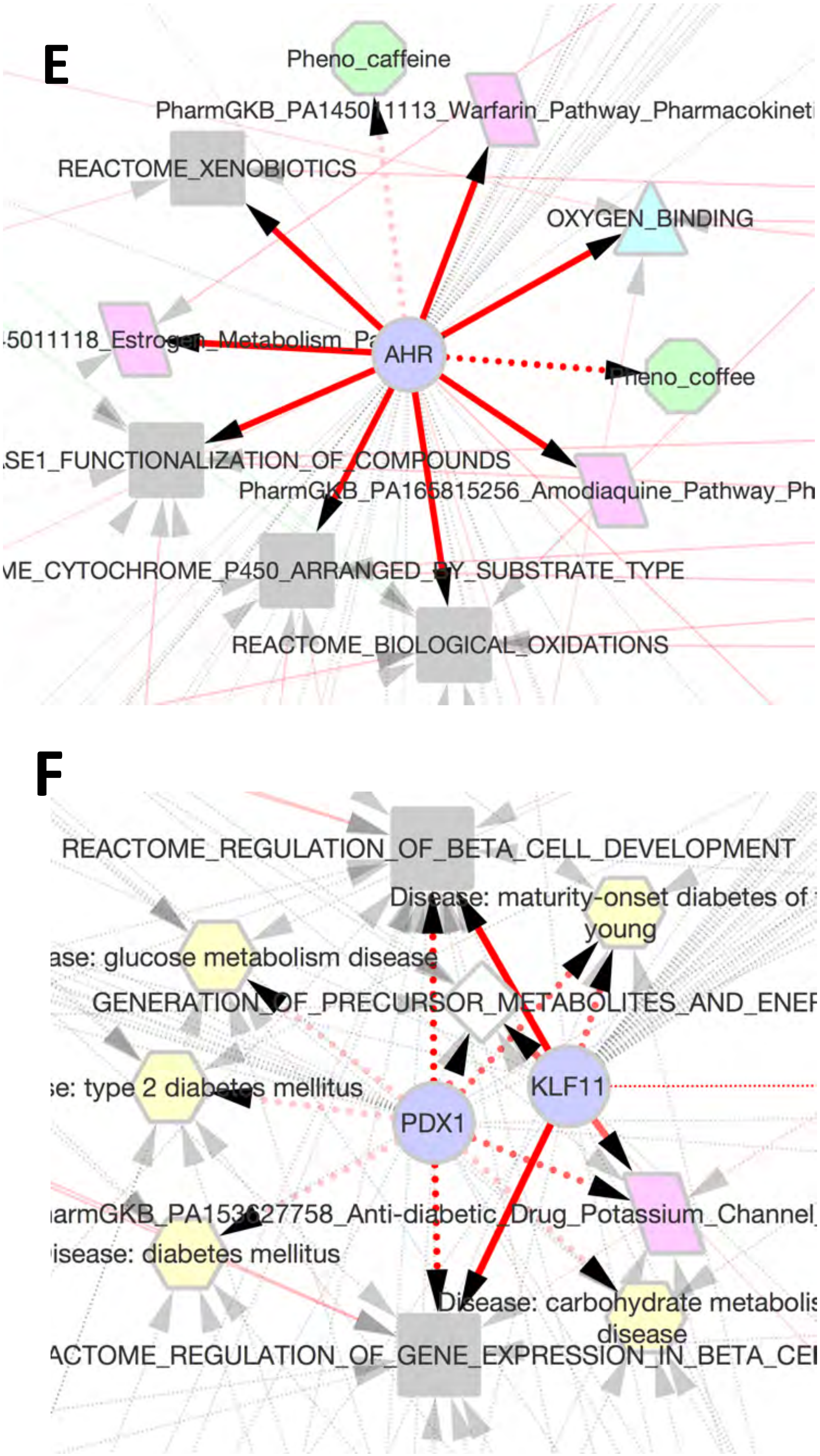

G

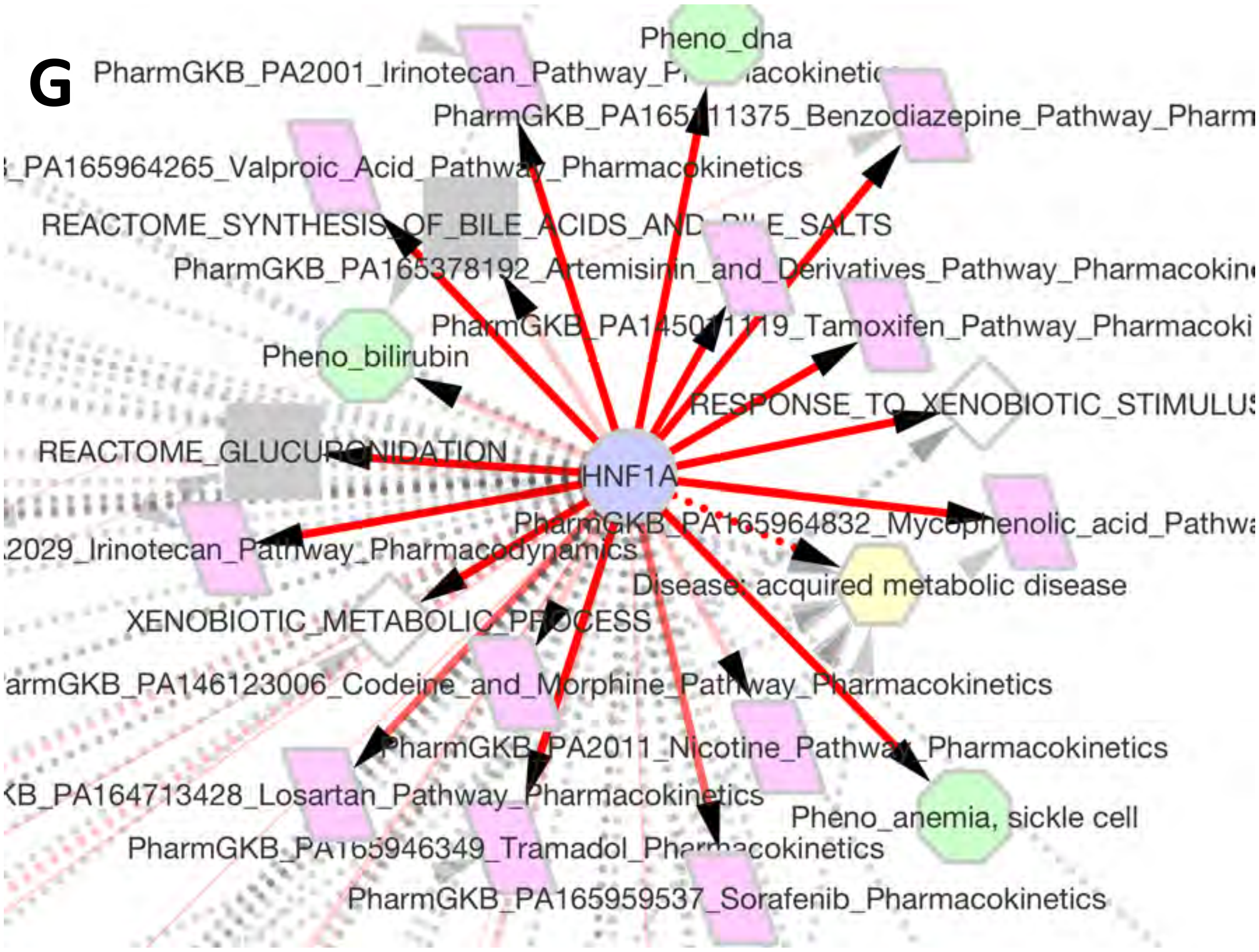
Pheno_dna
PharmGKB_PA2001_Irinotecan_Pathway_P
acokineti
PharmGKB_PA165
11375_Benzodiazepine_Pathway_Pharm
_PA165964265_Valproic_Acid_Pathw
_Pharmacokinetics
REACTOME_SYNTHESIS
OF_BILE_ACIDS_AN
E_SALTS
PharmGKB_PA165378192_Artemisinin_and_Derivatives_Pathway_Pharmacokin
PharmGKB_PA145011119_Tamoxifen_Pathway_Pharmacoki
Pheno_bilirubin
RESPONSE_TO_XENOBIOTIC_STIMULUS
REACTOME_GLUCURONIDATION
HNF1A
PharmGKB_PA165964832_Mycophenolic_acid_Pathwa
2029_Irinotecan_Pathway_Pharmacodynamics
Disease: acquired metabolic disease
XENOBIOTIC_METABOLIC_PROCESS
armGKB_PA146123006_Codeine_and_Morphine_Pathway_Pharmacokinetics
PharmGKB_PA2011_Nicotine_Pathway_Pharmacokinetics
B_PA164713428_Losartan_Pathway_Pharmacokinetics
Pheno_anemia, sickle cell
PharmGKB_PA165946349_Tramadol_Pharmacokinetics
PharmGKB_PA165959537_Sorafenib_Pharmacokinetics

Figure 4:

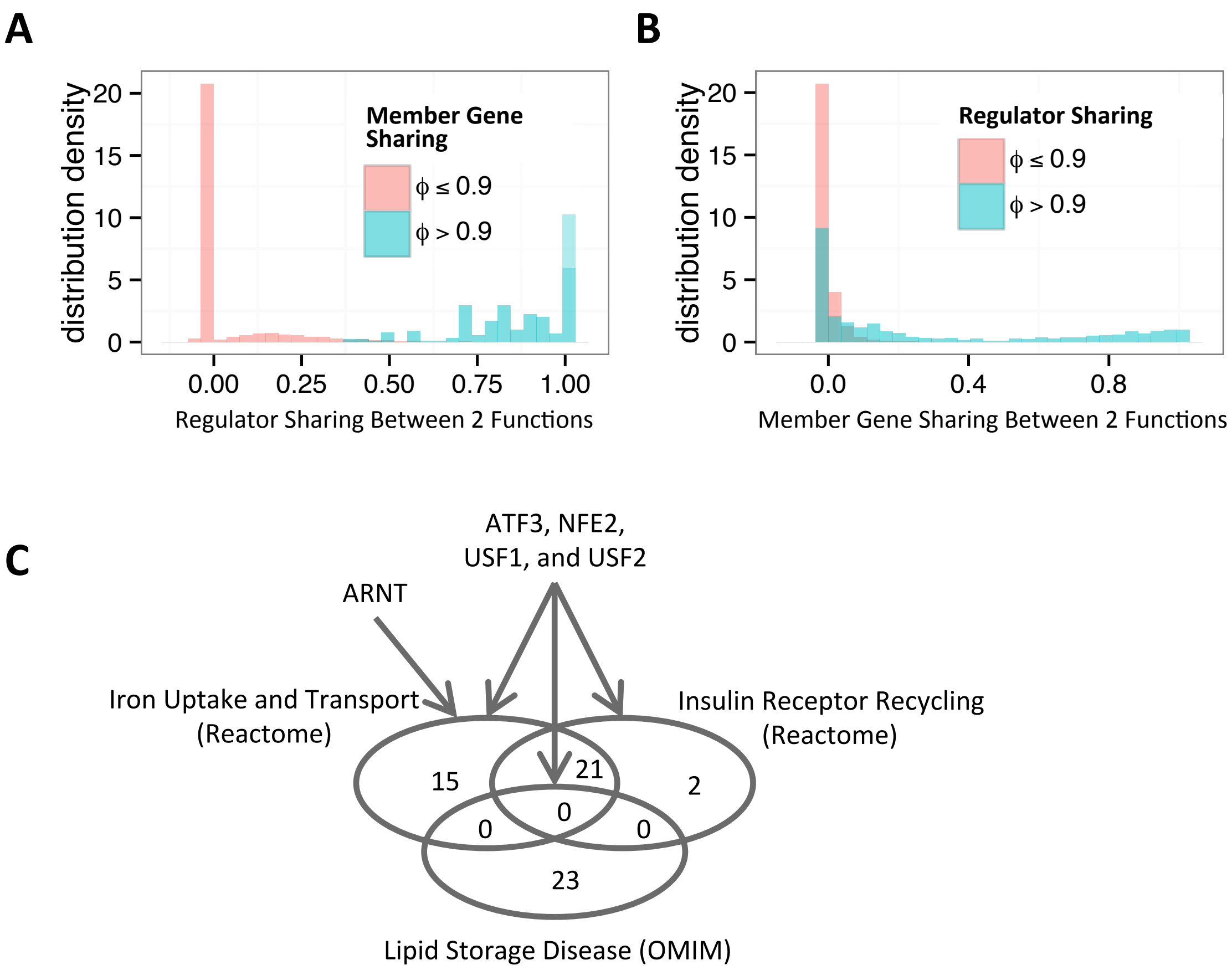

Figure 5:

A

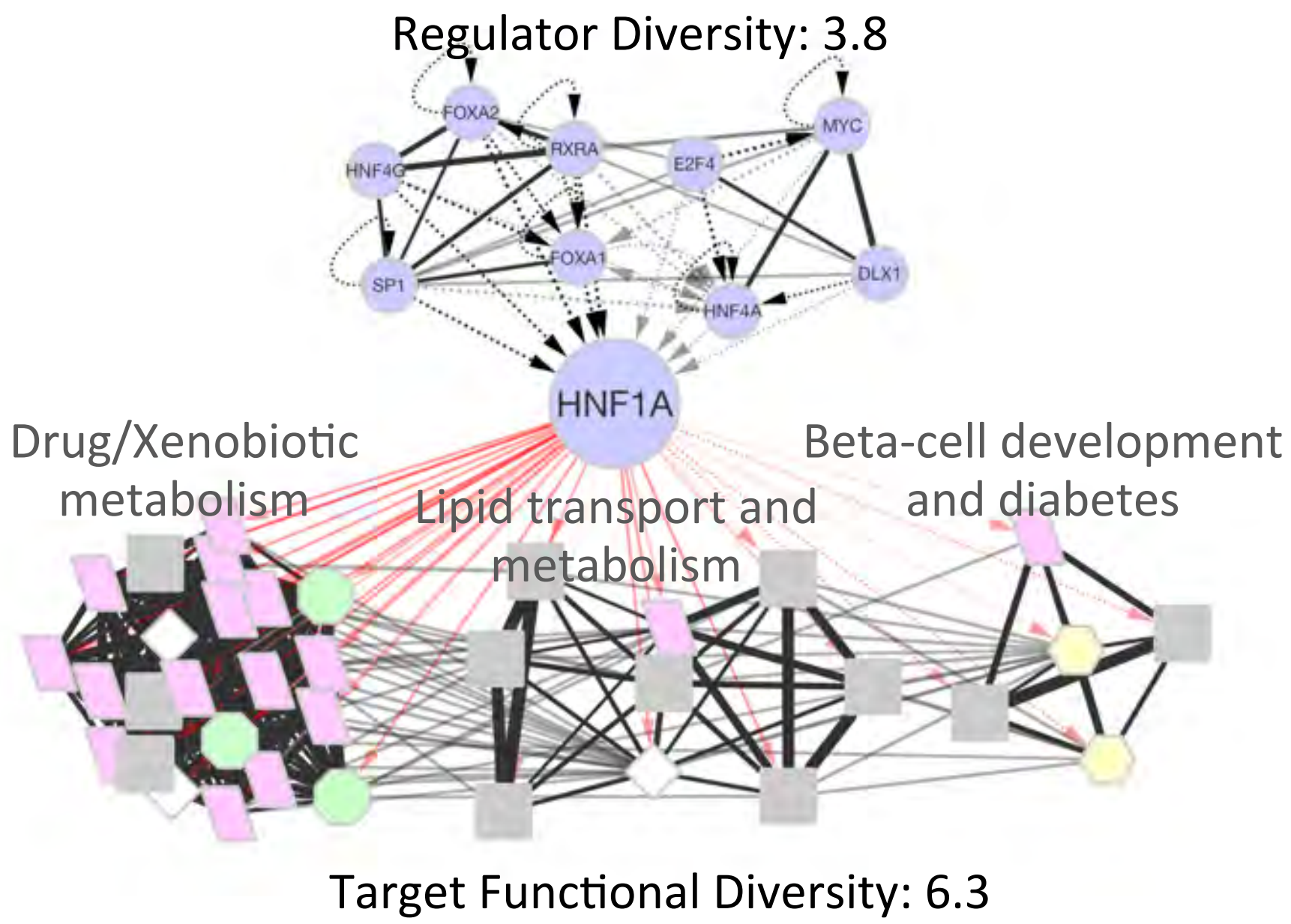

B

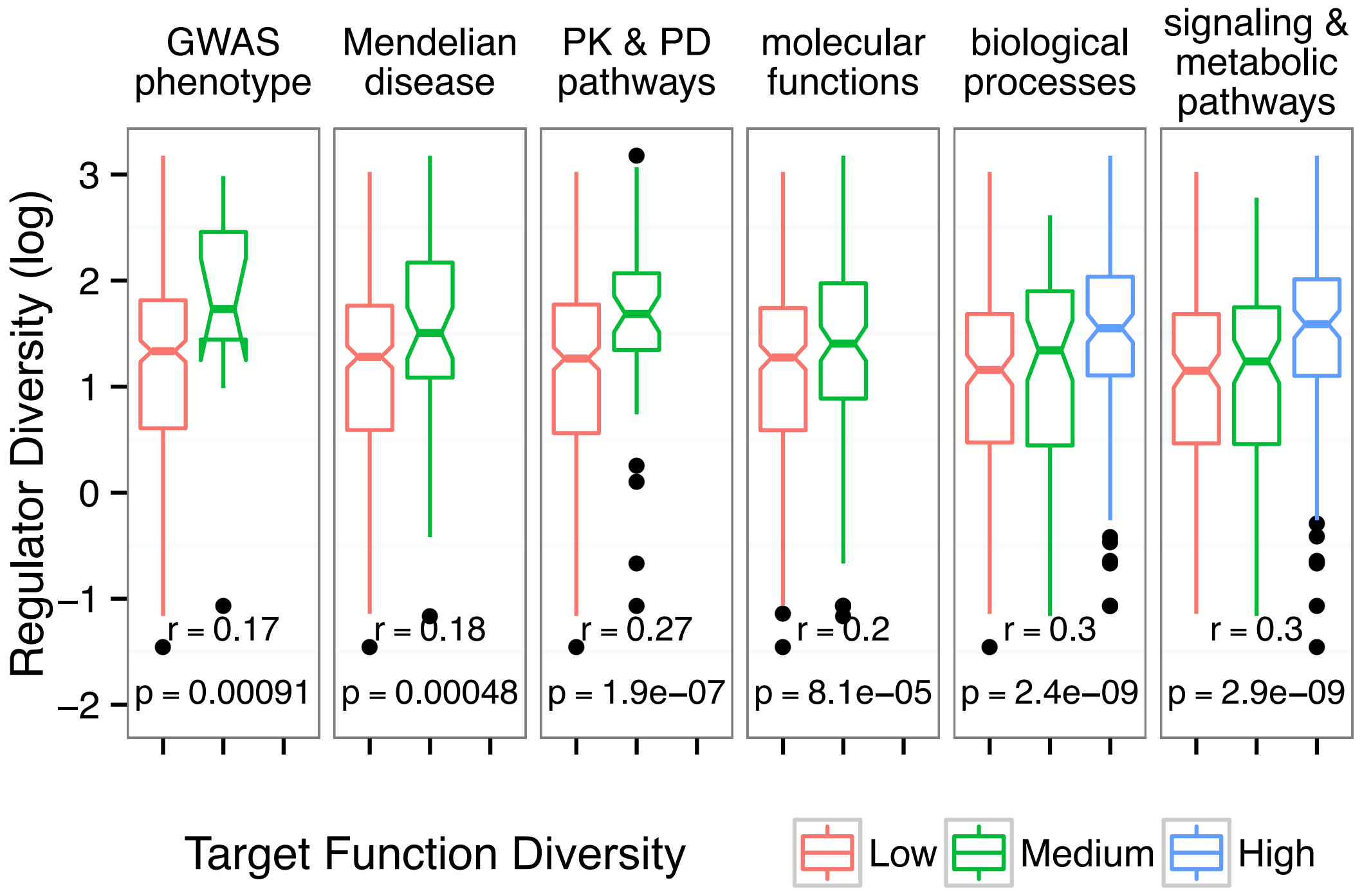

Figure 6:

**A**

| **Annotation of NFKB1** | **Similar TFs:** REL, BCL11A | |
|---|---|---|
| | **Significant Target Functions ($\log_{10}$ p-value)** | **Functional Diversity** |
| GWAS Phenotype | Arthritis, rheumatoid (8.4)<br>Psoriasis (6.95)<br>Colitis, ulcerative (6.29) | 2.0 |
| Mendelian Disease | Disease of cellular proliferation (7)<br>Organ system cancer (6.84)<br>Cancer (6.64)<br>+ 10 more | 2.5 |
| PD/PK Pathway | EGFR inhibitor pathway (PD) (6.13)<br>Doxorubicin pathway (cancer cell) (PD) (5.47) | 1.2 |
| Signaling/ metabolic Pathway | Immune system (17.11)<br>Cytokine signaling in immune system (14.62)<br>Interferon gamma signaling (10.89)<br>+ 46 more | 13.2 |
| Biological Process (GO) | Adaptive immune response (5.86)<br>Regulation of protein metabolic process (5.81)<br>Regulation of cytokine production (5.72)<br>+ 71 more | 10.2 |
| Molecular Function (GO) | Cytokine activity (14.33)<br>Receptor binding (8.3)<br>Chemokine activity (7.22)<br>+ 4 more | 3.2 |
| | **Total Diversity**: | 23.7 |

**B**

| **Annotation of SUZ12** | **Similar TFs:** CTBP2 | |
|---|---|---|
| | **Significant Target Functions ($\log_{10}$ p-value)** | **Functional Diversity** |
| GWAS Phenotype | - | 0 |
| Mendelian Disease | Heart septal defect (8.13)<br>Congenital heart disease (7.57)<br>Disease (7.14)<br>+ 6 more | 1.8 |
| PD/PK Pathway | - | 0 |
| Signaling/ metabolic Pathway | Regulation of beta cell development (12.19)<br>Regulation of gene expression in beta cells (6.26)<br>Class b 2 secretin family receptors (4.08) | 1.2 |
| Biological Process (GO) | Anatomical structure development (40.03)<br>Multicellular organismal development (33.22)<br>System development (32.56)<br>+ 29 more | 5.8 |
| Molecular Function (GO) | Transcription factor activity (36.68)<br>DNA binding (33.55)<br>RNA polymerase II transcription factor activity (11.83) + 1 more | 2.2 |
| | **Total Diversity:** | 8.3 |

# Tables

Table 1. Top twenty TF – target disease associations. The "Literature Rich" gene universe is used for the association detection. $Log_2$(OR), $log_2$ transformed odds ratio; P-value: p-value from single-tailed Fisher's exact test for odds ratio > 1; Evidence: lists published genetic evidence directly support the association of the TF with the disease; Mutation: mutations of the TF are observed in the disease or closely related diseases. Association: The TF gene locus is genetically associated with the disease or related diseases. Mouse: mouse model shows phenotypes directly related to the disease. Non-genetic evidence in the literature is not considered. MODY1: maturity-onset diabetes of the young. FCHL: familial combined hyperlipidemia.

| TF | Target Disease | $log_2$(OR) | P-value | Evidence |
|---|---|---|---|---|
| **ATF3** | Lysosomal storage disease | 4.5 | 3.9E-09 | - |
| **BRCA1** | Mitochondrial metabolism disease | 3.1 | 3.7E-09 | - |
| **CTBP2** | Heart septal defect | 6.9 | 4.4E-10 | Mouse (Hildebrand and Soriano 2002) |
| | Congenital heart disease | 6.5 | 1.6E-09 | Mouse (Hildebrand and Soriano 2002) |
| **ETS1** | Organ system cancer | 2.6 | 1.2E-08 | Mutation (Seth and Watson 2005) |
| **GATA1** | Acute porphyria | 7.4 | 9.8E-09 | Mutation(Phillips et al. 2007) |
| **HNF4A** | Mitochondrial metabolism disease | 3.0 | 5.3E-09 | Mutation in MODY1 (Wang et al. 2000) |
| **NFE2** | Lysosomal storage disease | 5.7 | 4.8E-12 | - |
| | Lipid storage disease | 6.0 | 5.8E-09 | - |
| **RFX2** | Bardet-Biedl syndrome | 5.5 | 1.1E-12 | Mouse (Bisgrove et al. 2012; Chung et al. 2012) |
| **SOX10** | Waardenburg's syndrome | 11.4 | 2.0E-09 | Mutation (Pingault et al. 1998) |
| **SUZ12** | Heart septal defect | 6.2 | 7.4E-09 | Mouse (He et al. 2012) |
| **TP53** | Organ system cancer<br>Cancer<br>Disease of cellular proliferation<br>Reproductive organ cancer | 3.5<br>3.5<br>3.4<br>4.8 | 1.5E-19<br>3.6E-19<br>1.7E-18<br>1.2E-08 | Mutations in multiple cancer Types (Malkin et al. 1990) |
| **USF1** | Disease of metabolism<br>Inherited metabolic disorder | 2.2<br>2.2 | 1.3E-10<br>7.5E-09 | Association with FCHL (Coon et al. 2005; Pajukanta et al. 2004) |
| **USF2** | Lysosomal storage disease<br>Disease of metabolism | 4.1<br>2.3 | 1.3E-09<br>2.8E-09 | -<br>- |

# 1. Supplemental Results

## 1.1. Negative Associations between Transcription Factors and Functions

Overall, 279 (73%) transcription factors are annotated by at least one functional term, while around 27% (103) transcription factors are annotated at least once by negatively associated gene functions or phenotypes. Percentage of negative TF association is lowest for pharmacogenomic pathways (2%) and Mendelian diseases (5%), and highest for GWAS phenotypes (51%) and gene molecular functions (40%). Many of the negative associations may simply indicate that the TFs and the function are irrelevant, although some of the negative TF-phenotype associations appear to be biologically meaningful. For example, NOTCH1 is negatively associated with myocardial infarction (OR = 0.47, Hypergeometric test p-value = $3.8 \times 10^{-7}$), while NOTCH1 is known to mediate cardiac repair following myocardial infarction (Gude et al. 2008; Li et al. 2011).

## 1.2. Transcription Factors have Preferences on the Type of Function Annotations

We observed extensive TF sharing among the 6 types of gene functions (**Fig. S5A**). Meanwhile, we observed systematic preference of different types of TFs to different sources of gene functions. For example, HNF1A target functions are 53 fold enriched for PK/PD pathways, 23 fold enriched for GWAS phenotypes, and 4 fold enriched for Mendelian diseases, while TCF12 target functions are 16 fold enriched for GO molecular functions, and 4 fold enriched for Reactome pathways. In total, there are 62 TFs with significant biases toward specific types of functions (**Fig. S5B)**. We note that different types of gene annotations are focused on different levels or different aspects of functions in living organisms. The biases of TFs toward specific types of functions suggest some high-level architecture of the TF-target function network. Specifically, different TFs may be in charge of functions of different levels from molecular to phenotype.

### 1.3. Transcription Factors Sharing Reveals Redundant or Related Functions

From the TF-target function relationships, we observed extensive regulator sharing among the functional concepts (e.g. *Cor pulmonale* and primary pulmonary hypertension) as well as target function overlaps among transcription factors (e.g. CTBP2 and SUZ12 in **Fig. 3B** top left). Redundancies or associations among functional concepts from both same sources and different annotation sources are observed (**Fig. 3B**). One cause of the concept similarities is the built-in redundancy of functional concepts within sources such as Gene Ontology, which defines concepts hierarchically, and between sources such as the GO biological processes and Reactome pathways, both of which covers signaling pathways. Another cause is the inherent biological relatedness of apparently different concepts, such as *beta-cell development* and *diabetes*, both of which are identified as targets of KLF11 and PDX1 (**Fig. 3B** bottom left).

We quantified the redundancy between functional concepts by their member gene overlap using Pearson's $\phi$ coefficient, and estimated the total effective number of functional concepts to be 1316 compared to the total of 3715. Relatedly, 954 pairs of functional concepts are regulated by identical set of TFs, and 141 pairs of concepts have identical member genes. GO biological processes have the highest redundancy of 3.7 fold, while the GWAS phenotypes have the lowest redundancies of 1.2 fold (**Table S2**).

To further reveal the overall concept relatedness/redundancy between different sources of functional concepts, we calculated the average number of shared TFs per pairs of functional concepts from the 6 different sources (**Fig. S5B**). We find molecular functions, biological processes, and Reactome pathways cluster closely with >0.5 TFs shared per pair of functions from different sources, compared to 0.6-1.3 TFs shared per pair of functions from same sources). Pharmacogenomic pathways and Mendelian diseases form another cluster with 0.08 TFs shared per pair of functions from different sources, compared to 0.16 or 0.22 TFs shared per pair of functions from same sources.

## 1.4. Basic Statistics of Functional Diversity and Regulator Diversity of TFs

Overall, the average uniqueness of all functional concepts is only 0.4 (i.e. average degeneracy 2.8). There are on average 23.9 target functions per TF, while the target function diversity is only 3.7 with a degeneracy of around 6.1 fold, higher than the average degeneracy of all functions. On the other hand, the transcription factors are also highly related to each other, with an average uniqueness of 0.6, and effectively 226.0 functionally unique TFs out of 384. For each TF, there are 11.8 upstream regulators, while the regulator diversity is 5.2 and regulator degeneracy 2.2. **Fig. 6A** give an example of TF HNF1A, which have 6.3 effective target functions and 3.8 effective regulators (see **Fig. S11** for more transcription factors RXF5 and NFKB1; and also examples of non-TF gene MTHFR).

### 1.5. Removing the impact of human research biases on the association between TF regulator diversity and function diversity

The TFTG compendium contains TF-target gene relationships from the literature of low throughput studies. Such data are prone to human biases toward the perceived important TFs. These popular TFs can have more target genes as well as upstream regulators reported due to higher research efforts, hence creating an artificial association between the number of regulators and number of identified target functions. We corrected such biases by including the number of target genes of TF as a confounding variable (see main text). As a stricter validation, we completely removed the low-throughput data in the TFTG data, retained only the target genes for 262 TFs from ChIP-seq studies, and redid all analyses. The selection of TFs in ChIP-seq experiment remains biased toward the well-known TFs. Popular TFs however should no longer have more upstream TFs due to preferred study of the regulators of these TFs. Despite a smaller number of TFs and a loss of the most reliable target genes for the remaining TFs, regulator diversity and function diversity remain significantly associated (p-value 0.015, or 0.0086 when controlling for the number of target genes).

## 1.6. Hierarchical location of a TF in the regulatory network is not associated with multifunctionality of the TF

Aside from regulation diversity, we discover that the hierarchical seniority of a TF, as measured by the PageRank of a TF in the core TRTG network, is also associated with the (known) function diversity (p-value 0.0023, PCC 0.15) and the regulator diversity of a TF (p-value 0.025, PCC 0.12). There is an even stronger association between PageRank and the target function diversity of a TF (p-value 1.2e-42 and PCC 0.62). These indicate that a transcription factor at the top of the regulatory hierarchy tends to have more diverse functions. This correlation however may be caused by the shared association of PageRank and the target-function diversity with the number of target genes of a TF, as they are both defined on the target genes. Indeed, after controlling for the target gene size through a linear model, neither the PageRank-function diversity or PageRank-target function diversity associations are significant (p-values 0.94 and 0.051 respectively). Similar behavior is observed for other TF hierarchical rank measures defined based on breadth first search (BFS) and percentage of regulatory target genes (Yu and Gerstein 2006; Bhardwaj et al. 2010).

## 1.7. Gene's Regulator Diversity and Multifunctionality

Significant associations between function and regulator diversities of genes are observed for specific types of functional annotations (**Supplemental Fig. S10**), including signaling/metabolic pathways (p-value $1.5 \times 10^{-6}$), biological processes (p-value $1.5 \times 10^{-6}$), and GWAS phenotypes (p-value $1.1 \times 10^{-10}$), while the diversity of GWAS phenotype of a gene is inversely correlated with its regulator diversity. Genes regulated by 10 or more other TFs are 14% less likely to be associated with any GWAS phenotypes, and 46% less likely to be associated with 5 or more phenotypes. Note that we used the Coding gene universe when analyzing GWAS phenotype associations, and the Literature Rich gene universe when analyzing other annotations. Both of these are conservative choices of gene universes. Using Literature Rich gene universe will lead to more significant negative association between TF regulation and phenotypes, while using Coding gene universe will lead to more significant positive associations between TF regulation and signaling & metabolic pathways, biological processes, and molecular functions.

We removed the low-throughput data in the TFTG data, and only retained the target genes for 262 TFs from ChIP-seq studies. The associations of TFs' regulator diversity with specific types of functions are PCC 0.17, p-value 0.007 for GWAS phenotypes; PCC -0.0063, p-value 0.92 for Mendelian disease; PCC 0.11 p-value 0.093 for PK & PD pathways; PCC 0.086, p-value 0.17 for molecular functions, PCC 0.14, p-value 0.032 for biological processes; PCC 0.15, p-value 0.015 for signaling & metabolic pathways. Meanwhile, for general genes (TFs and non-TFs), there is still a significant positive association between regulator diversity and signaling/metabolic pathway diversity (PCC 0.072, p-value 1.3e-6), and a significant negative association for GWAS phenotypes (PCC -0.11, p-value 6.8 e-19).

For the 10792 general genes that have regulators, function annotations as well as expression data, genes' regulator diversity and expression diversity are correlated (Spearman rank correlation 0.22, p-value $3.7 \times 10^{-120}$ for diversities, or correlation 0.26, p-value $1.0 \times 10^{-165}$ for the raw counts), and genes' regulator diversity and function diversity are also correlated (Spearman rank correlation 0.12, p-value $1.0 \times 10^{-33}$).

## 2. Supplemental Methods

### 2.1. Modified TIP algorithm for Exacting Target Genes from ChIP-Seq Data

The binding signals from target genes were differentiated from that from non-target genes using a modified version of the TIP algorithm (Cheng et al. 2011). Specifically, the modified implementation accepts both fixed step and variable step wiggle formatted files as well as other commonly used genome feature formats, such as bed and bigWig. It properly handles the 0-based start and 1-based end coordinates of transcript annotations and the 1-based coordinates in wig files. Genomic regions with read density less than 1 read per million total reads are ignored. In addition, it only evaluates the unambiguously mapped transcripts, and removes redundant promoter regions (i.e. transcripts with identical TSS are merged). RefSeq gene annotations on human genome hg18 and hg19 were obtained from the UCSC genome browser (https://genome.ucsc.edu) (Karolchik 2003). Instead of the dot product scoring functions used in TIP, we used log likelihood ratio to achieve a Gaussian like distribution of the scores,

$$S_g = \sum_{l=-3000}^{3000} x_{gl} \cdot \log\left(\frac{m_{l.gene}}{m_{l.null}}\right),$$

where $x_{gl}$ is the read counts at location $l$ within the promoter region, and score $S_g$ for promoter/gene $g$ is calculated as the read-counts weighted sum of the log likelihood ratio between the average signal distribution $m_{l.gene}$ and the even distribution $m_{l.null}$. A 6k window is used spanning ±3K base pairs 5’ and 3’ to the TSS. P-values are then computed by assuming a Gaussian distribution of $S_g$. The target genes of each TF in each experiment are then obtained at q-value cutoff of 0.01 (Storey 2003; Storey and Tibshirani 2003). Multiple experiments or data sources for the same TF are then merged, resulting in target gene sets for 384 unique transcription factors.

## 3. Supplemental Figures

Figure S1:

A. Transcription factor coverage by three sources of TF-target gene data. LTP: low throughput experiments; ENCODE: TF ChIP-seq experiments in ENCODE study; Other ChIP-seq: non-ENCODE ChIP-seq experiments compiled from GEO.

B-E. Transcription factor (TF) and target gene (TG) degree distributions in the TFTG data compendium and the subset of TF-target TF core network. (B) TG degree (i.e. node in-degree) and (C) TF degree (i.e. node out-degree) distributions in the full TFTG data compendium. (D) TG degree (i.e. node in-degree) and (E) TF degree (i.e. node out-degree) distributions in the core TRTG data compendium for which the TGs are restricted to be only the 384 TFs.

Figure S2:

Venn diagram showing the overlaps among five types of gene function annotations. Mendelian diseases are not included in the Venn diagram due to limitation of the visualization technique.

Figure S3:

The significance of TF-target function association depends on the choice of gene universe. Shown here is the impact of gene universe on the association between TF SP1 and Reactome pathway *Immune System*. Three gene universes are evaluated, while the literature rich gene universe is the one used in this study.

Figure S4:

A. The in-degree and out-degree distributions of the TF-Target Function network follow a power law distribution with exponents 1.2 approximately. Left: distribution of the out-degrees of TFs (#Functions per TF). Right: distribution of the in-degrees of functions (#TFs per function).

B. The TFs that are not annotated with any target functions (blue curve) have less target genes than TFs that are annotated with one or more target functions.

Figure S5:

A. Average number of transcription factors shared per pair of function annotations. FDR cutoff 0.05 was applied to select significant target functions. The clustering is based on 1 – diagonal-normalized TF sharing, which is 1-average(cosine similarity) of regulator vectors of functional concepts.

B. Shown are heatmap of 62 TFs with significant annotation type biases at q-value cutoff 0.01. Clustering are based on cosine similarity of the counts of significant TF-Target function associations. The color in the heatmap correspond to log(Cij*C++/Ci+*C+j), where Cij is the counts for TF i and function type j, C++ is the total count, Ci+ is the total number of significant target functions for TF i, and C+j is the total number of significant TFs for a function type j. Positive values (red) indicate favored function types, and negative values (all set to -1, green color) indicate disfavored function types.

Figure S6:

Receiver Operator Characteristics (ROC) curves measuring the predictive performances of TF-Target function associations (phi) against known TF-function annotations. AUC, area under the ROC curve; FET, Fisher's Exact Test; Centered p-value, 2 times the min of the two single tailed FET p-values.

Figure S7:

Clustering of TFs by target gene-based TF-TF similarity (measured by Pearson's phi). In the heatmap, colors green to red correspond to negative to positive TF-TF target gene associations. Hierarchical clustering of TFs is performed using 1-phi as distance measure. Six selected sub-trees in the clustering dendrogram are shown (see **Figure S8** for a larger version of the full dendrogram).

Figure S8:

Evolutionary and functional trees of 384 TFs covered by the TFTG data compendium.

(A) Protein sequence based phylogenetic trees of TFs. Sequence similarity based clustering of transcription factors using UPGMA.

(B) Target-gene similarity (phi) based functional clustering of transcription factors, same as the clustering shown in Figure 6.

Figure S9:

Functional concepts that share TFs do not necessarily share member genes significantly.

(A) Two target functional concepts with 100% member gene overlaps always have identical regulators, but (B) two functional concepts with 100% identical regulators do not always have high member gene sharing.

Figure S10:

Significant association exists between the regulator diversity and functional diversity of 11345 genes that have both regulators and function annotations.

(A) Six types of function annotations are analyzed separately and significant positive association is observed for biological processes and signaling/metabolic pathways.

(B) Estimated odds ratio for tightly regulated genes (number of TF >= 10) versus annotated genes (number of annotation > 0).

Figure S11:

The upstream regulators and downstream (target) functions of TFs (A) NFKB1 and (B) RFX5, and non-TF gene MTHFR. The coloring schema is same as in Figure 5 and the clustering of TFs and functions are based on the target gene and member gene overlaps.

Figure S12:

Transcription factor's regulator diversity is associated with target function diversity. Pearson correlation is 0.32, p-value 3.3e-10; and for the TF with over 100 target genes, correlation is 0.35, p-value 6.0e-06. HNF1A is highlighted in red.

Figure S1:

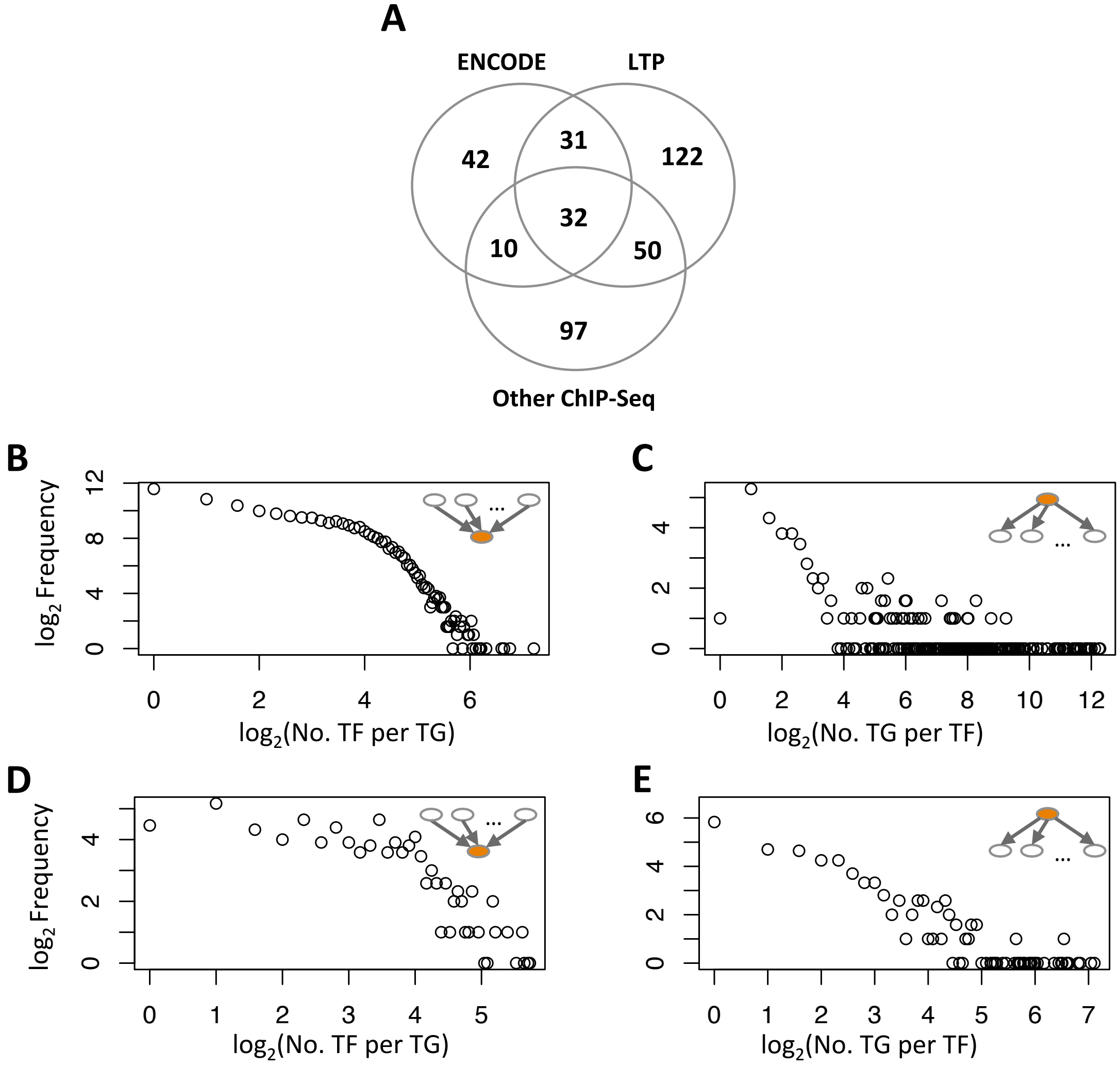

Figure S2:

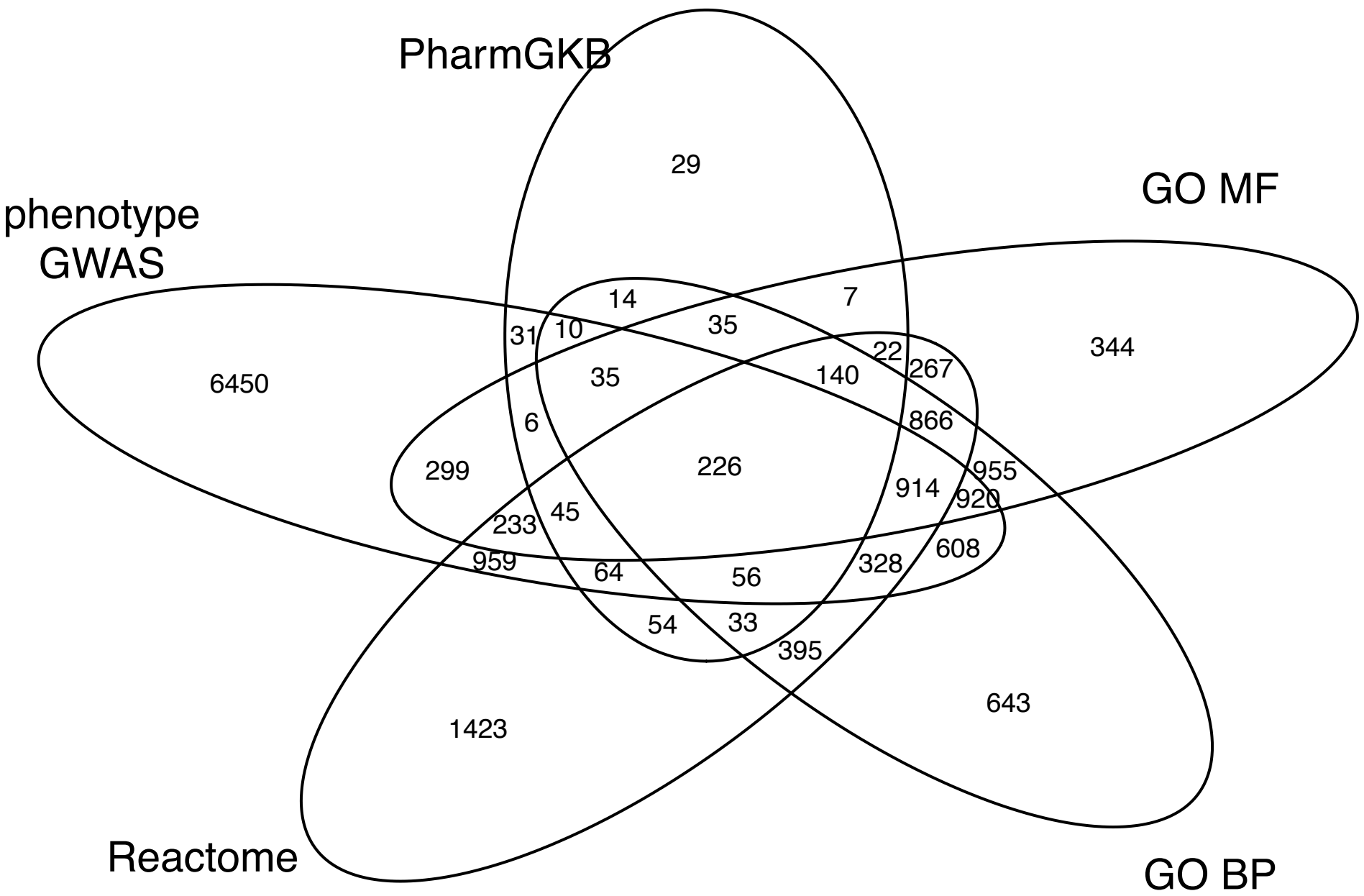

Figure S3:

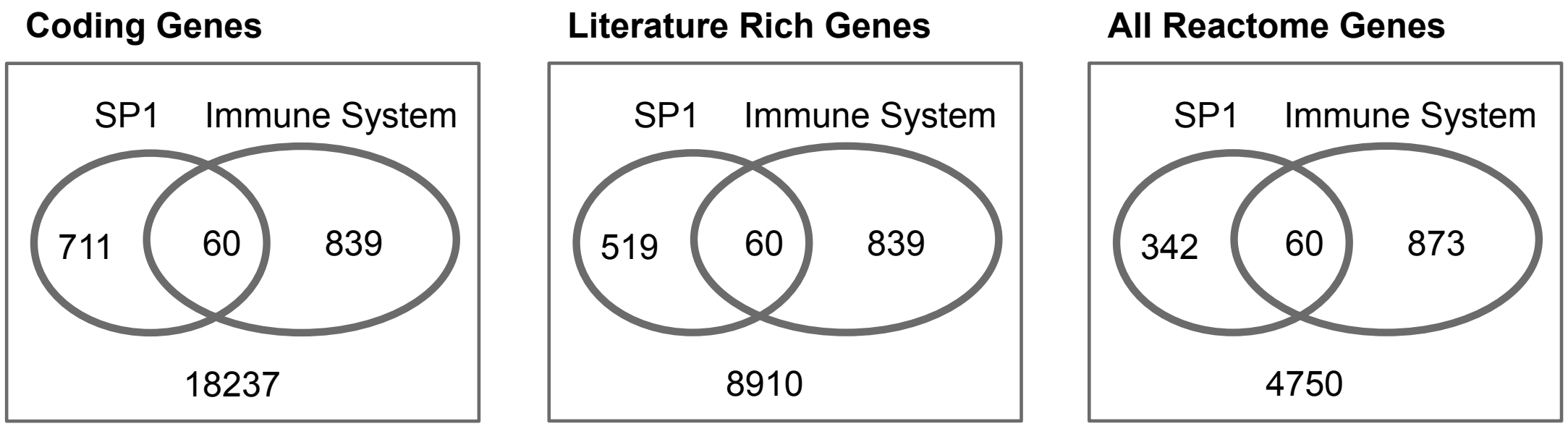

Positive association:
p-value = $3.2 \times 10^{-5}$; odds ratio = 1.9; FDR = $2.8 \times 10^{-3}$

Positive association:
p-value = 0.086; odds ratio = 1.2
FDR > 0.1

Negative association:
p-value = 0.41; odds ratio = 0.97
FDR > 0.1

Figure S4

A

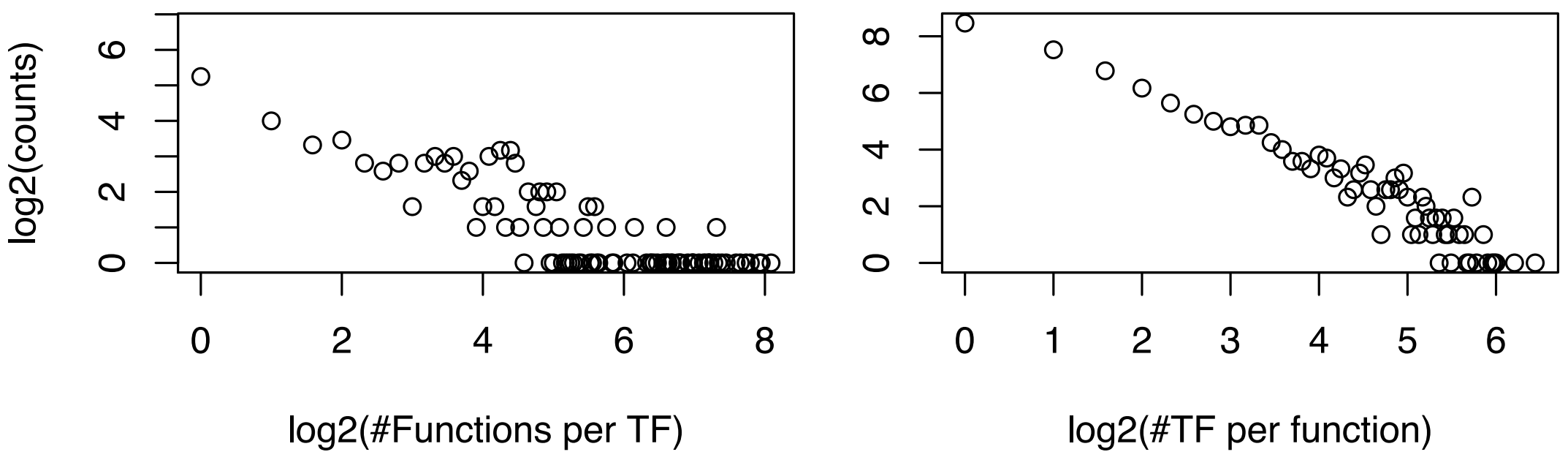

B

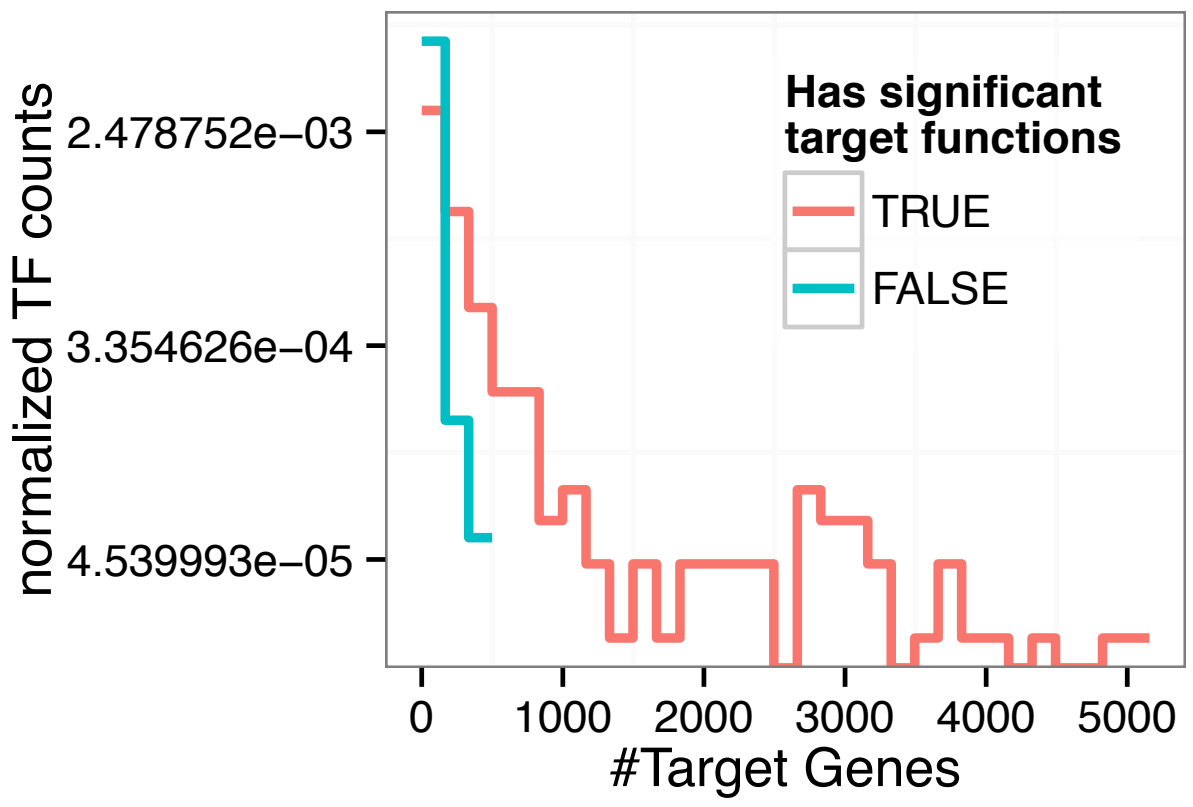

Figure S5:

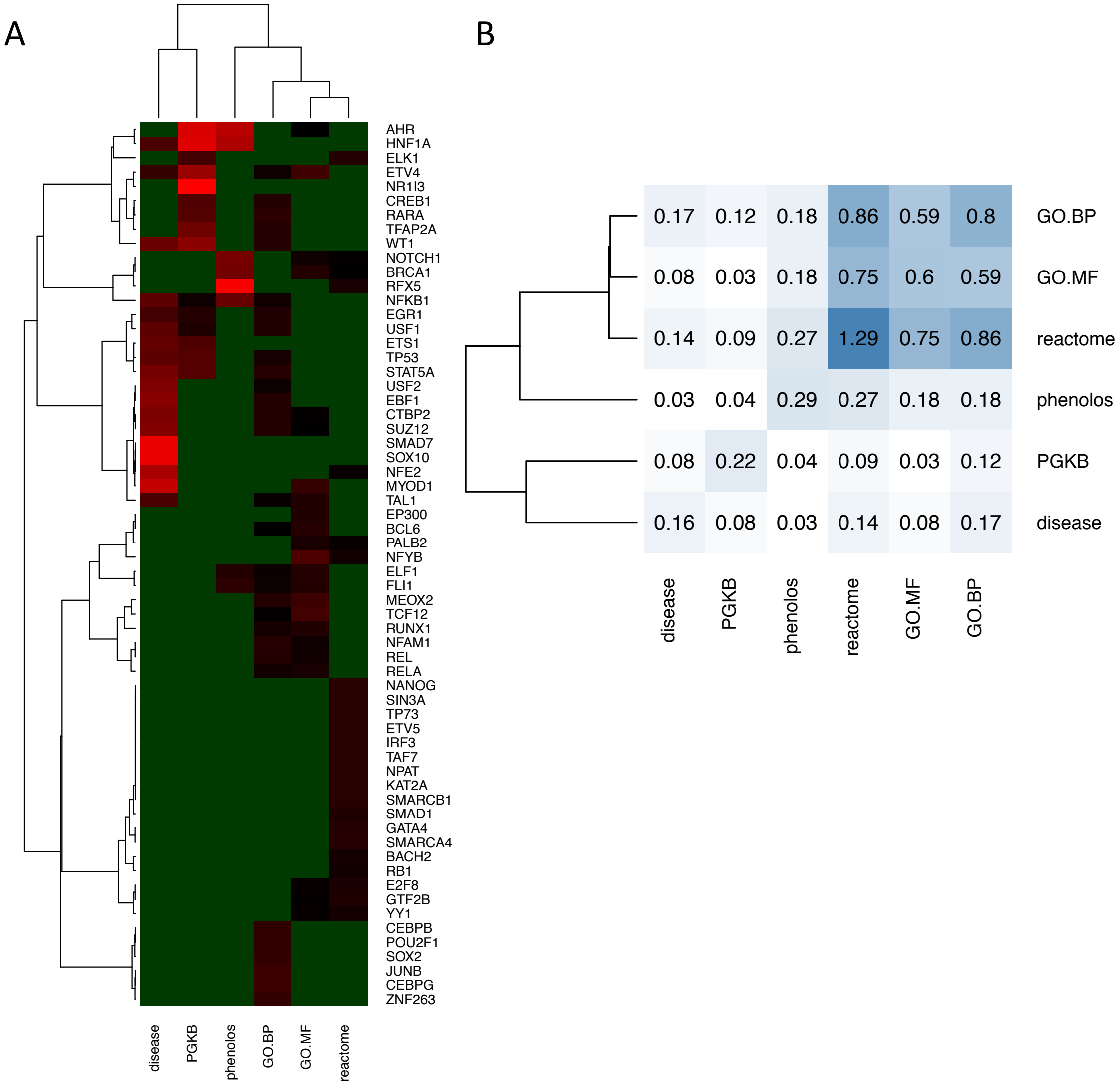

Figure S6:

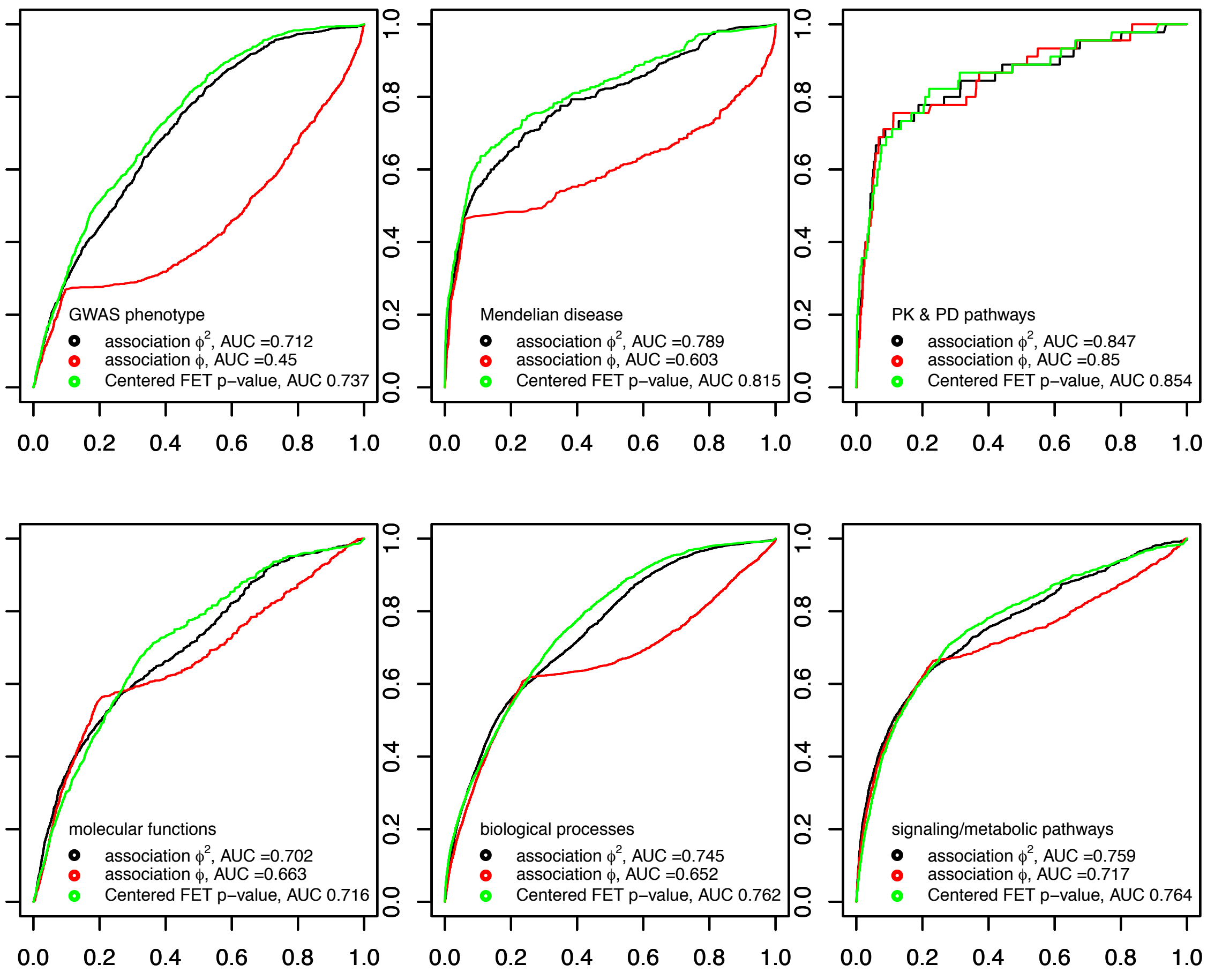
GWAS phenotype
association $\phi^2$, AUC =0.712
association $\phi$, AUC =0.45
Centered FET p−value, AUC 0.737
Mendelian disease
association $\phi^2$, AUC =0.789
association $\phi$, AUC =0.603
Centered FET p−value, AUC 0.815
PK & PD pathways
association $\phi^2$, AUC =0.847
association $\phi$, AUC =0.85
Centered FET p−value, AUC 0.854
molecular functions
association $\phi^2$, AUC =0.702
association $\phi$, AUC =0.663
Centered FET p−value, AUC 0.716
biological processes
association $\phi^2$, AUC =0.745
association $\phi$, AUC =0.652
Centered FET p−value, AUC 0.762
signaling/metabolic pathways
association $\phi^2$, AUC =0.759
association $\phi$, AUC =0.717
Centered FET p−value, AUC 0.764
0.0
0.2
0.4
0.6
0.8
1.0

Figure S7:

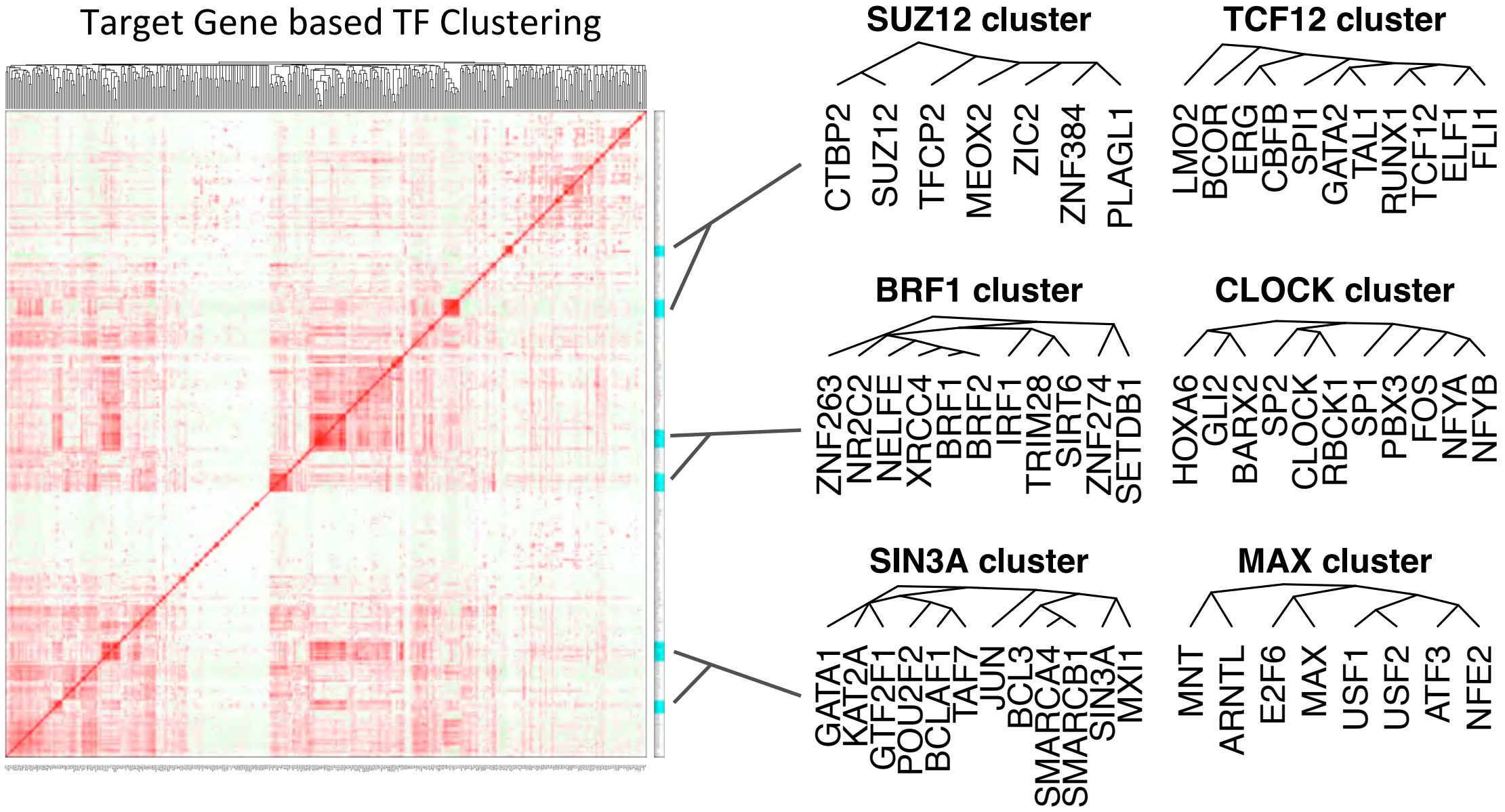

Figure S8:

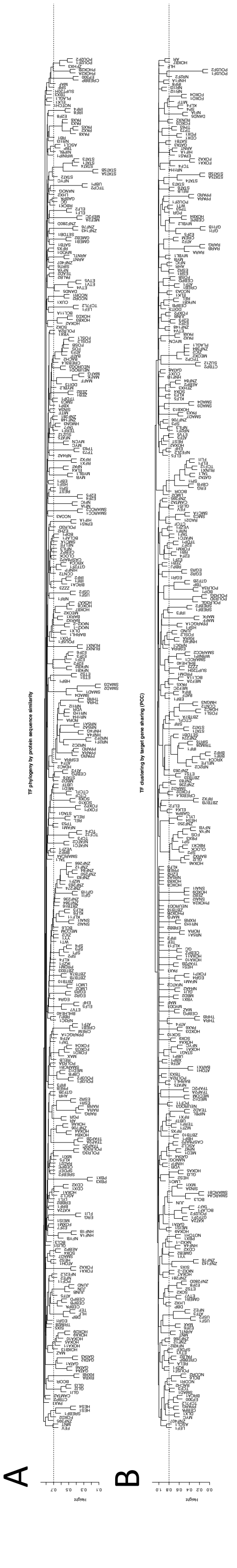

A
B
TF phylogeny by protein sequence similarity
TF clustering by target gene sharing (PCC)
Height

Figure S9:

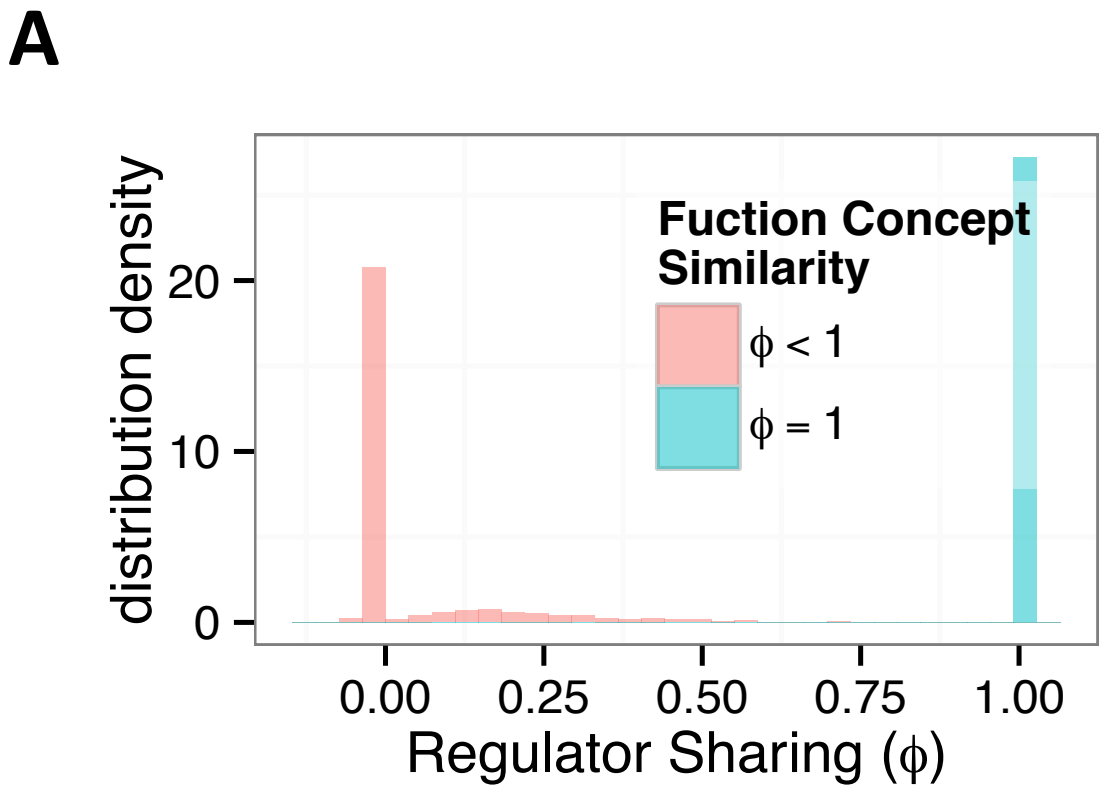

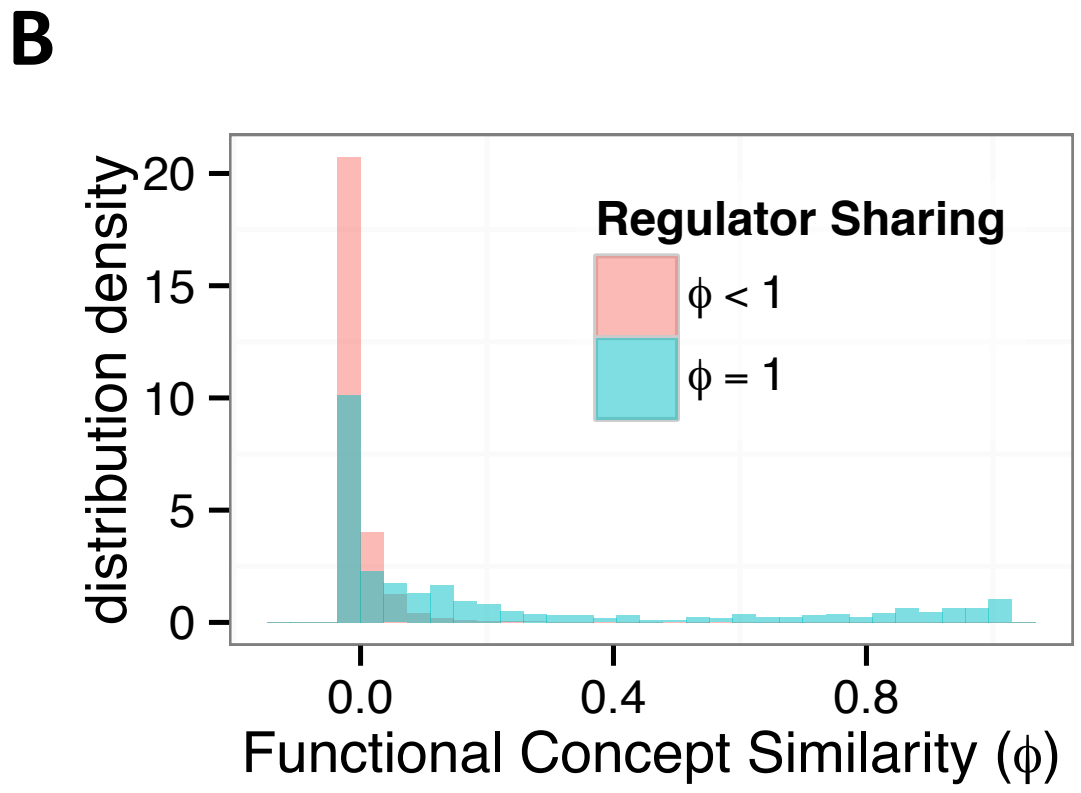

Figure S10:

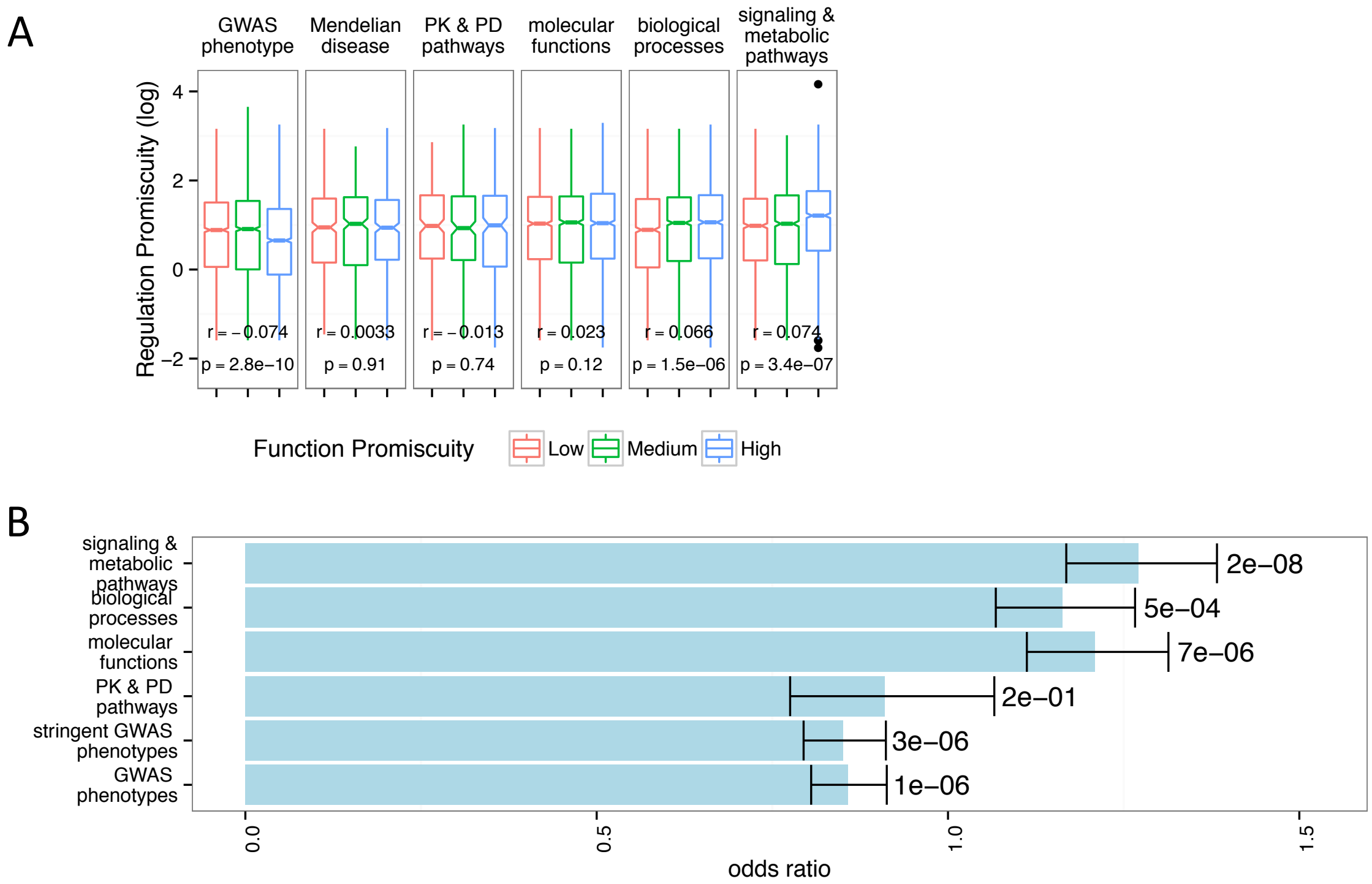

Figure S11:

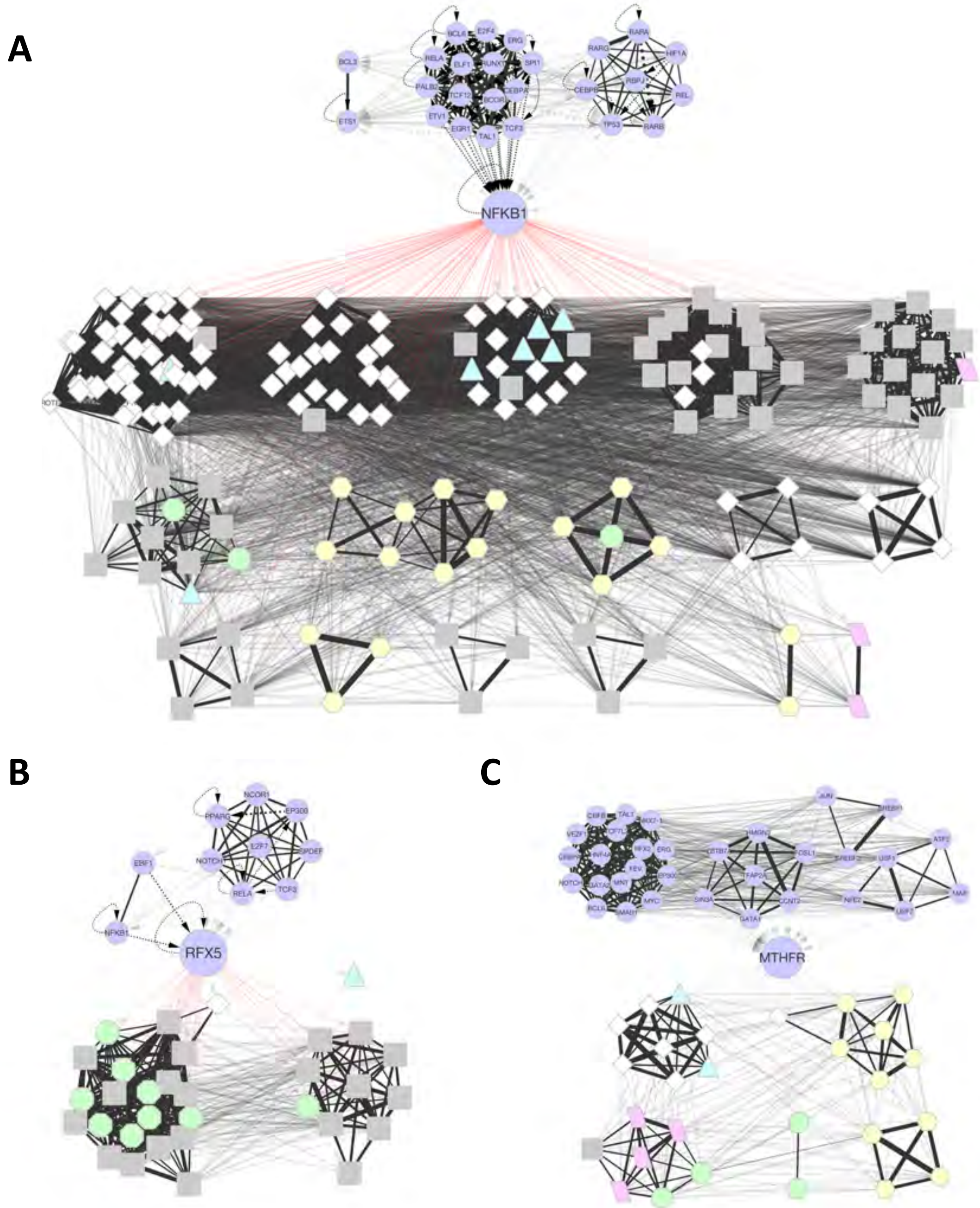

Figure S12:

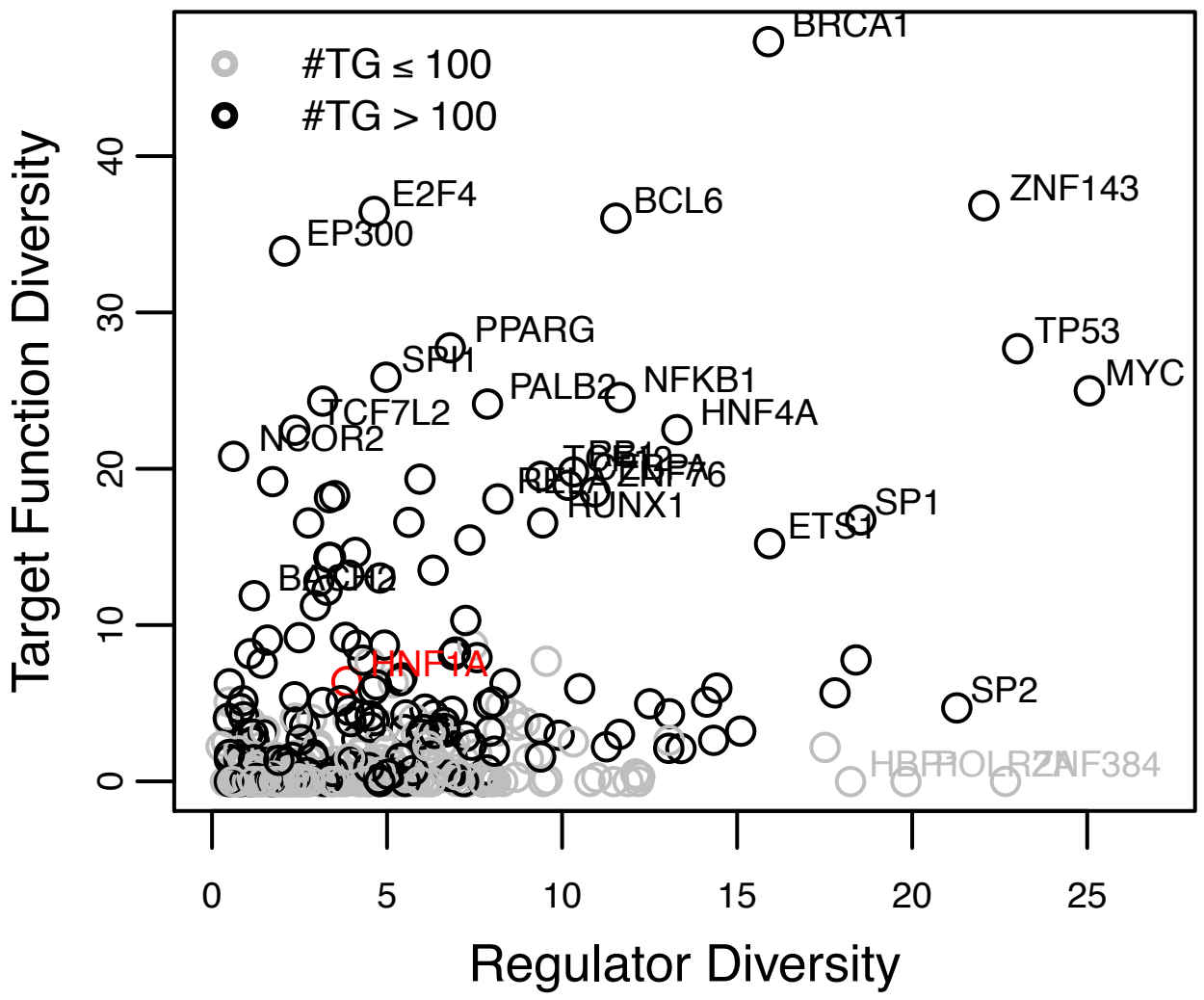

## 4. Supplemental Tables

Table S1: A summary of the total number and effective number of concepts in each of the 6 function annotation databases. Phenotype: complex phenotypes studied in GWAS; Diseases: Mendelian Diseases; PharmGKB: PK & PD pathways from PharmGKB; MF: Molecular Functions from GO; BP: Biological processes from GO; Pathways: Signaling & Metabolic Pathways from Reactome.

| | Phenotype | Disease | PharmGKB | MF | BP | Reactome |
|---|---|---|---|---|---|---|
| Functional Concept Counts | 573 | 1156 | 91 | 396 | 825 | 674 |
| Within DB Redundancy-Adjusted Counts | 467.7 | 551 | 44.9 | 176.8 | 225.2 | 217.3 |
| Between DB Redundancy-Adjusted Counts | 368.4 | 465.3 | 24.8 | 115.3 | 178.7 | 163.3 |

Table S2. Average number of significant functional concept enrichments per transcription factors at FDR 0.05. Coding gene universe is used for phenotypes and literature rich gene universe for the rest.

| | Phenotype | Disease | PharmGKB | MF | BP | Reactome |
|---|---|---|---|---|---|---|
| **# Enrichment** | 0.052 | 0.471 | 0.263 | 1.383 | 7.992 | 11.216 |

Table S3: TFs and their target complex phenotypes at FDR 0.05. The Coding gene universe is used for the association analysis. Only enrichment (positive associations) and the corresponding single tailed p-values are shown. log2(OR): log2 transformed odds ratio. Evidence: lists published genetic evidence directly supports the association of the TF with the disease; Mutation: mutations of the TF are observed in the disease or closely related diseases. Association: The TF gene locus is genetically associated with the disease or related diseases. Mouse: mouse model shows phenotypes directly related to the disease. Non-genetic evidence in the literature is not considered. # MHC II: RFX5 is mutated in Bare lymphocyte syndrome II (DeSandro et al. 1999; Reith and Mach 2001), and RFX5 knock out mouse shows abnormality in the immune system (Masternak et al. 1998; Clausen et al. 1998)

| TF | Phenotype | $log_2$(OR) | P-value | Evidence |
|---|---|---|---|---|
| **AHR** | Coffee consumption | 12.5 | 2.3E-07 | Association (Sulem et al. 2011; Cornelis et al. 2011, 2015) |
| | Habitual caffeine consumption | 10.1 | 5.0E-06 | Association (Sulem et al. 2011; Cornelis et al. 2011, 2015) |
| **HNF1A** | Serum cell-free DNA (Cardiovascular Risk) | 8.9 | 1.3E-09 | Association (Reiner et al. 2009), Mutation (Steele et al. 2010) |
| | Sickle-cell anemia (F-cell levels) | 7.6 | 5.6E-10 | - |
| | Serum bilirubin levels (bilirubin and cholelithiasis risk) | 5.7 | 3.8E-06 | Mouse (Pontoglio et al. 1996) |
| **NFKB1** | Psoriasis | 3.6 | 1.1E-07 | - |
| | Ulcerative colitis | 3.3 | 5.1E-07 | Association (Jostins et al. 2012) |
| | Rheumatoid arthritis | 3.2 | 4.0E-09 | - |
| **POU2F2** | Celiac disease | 4.2 | 4.7E-06 | - |
| **RFX5** | Chronic hepatitis b | 7.1 | 8.2E-06 | - |
| | Nasopharyngeal neoplasms | 6.5 | 9.5E-07 | - |
| | Systemic scleroderma | 6.1 | 2.8E-10 | MHC II # |
| | Leprosy | 4.7 | 7.5E-10 | - |
| | Psoriasis | 3.7 | 4.9E-06 | MHC II # |
| | Type 1 diabetes mellitus | 3.7 | 1.9E-09 | MHC II # |
| | Multiple sclerosis | 3.3 | 2.4E-08 | MHC II # |
| | Behcet syndrome | 3.3 | 8.8E-06 | MHC II # |
| | Systemic lupus erythematosus | 3.2 | 1.0E-06 | MHC II # |
| **SREBF1** | Low density lipoproteins (LDL) | 3.9 | 2.2E-06 | Mouse (Shimano et al. 1997) |
| **TBP** | Systemic scleroderma | 4.8 | 5.9E-06 | - |

Table S4: TF and their target pharmacogenomic pathways at FDR 0.05. The Literature Rich gene universe is used for the association detection. $Log_2$(OR), $log_2$ transformed odds ratio; P-value: p-value from single-tailed Fisher's exact test for odds ratio > 1; Evidence: For the PD pathways, we listed direct evidence supporting the TF-PD pathway relationship if, 1) Member: the TF is annotated as a member gene in same or closely related PharmGKB/Reactome pathways or GO

biological processes; 2) Census or Mutation: the TF is a known causal gene of the diseases based on Cancer Gene Census (Futreal et al. 2004) or OMIM (Hamosh et al. 2005); 3) GWAS: the TF gene locus is strongly associated with related phenotypes in GWAS. 4) Mouse: mouse model shows phenotypes directly related to the disease. Other forms of evidence in the literature are not considered. * PD pathways containing 50% or more PK genes. & neural normal: this is not disease related, but a normal pathway in neurological processes. N/A: not applicable.

| TF | Pharmacogenomics pathways | $\log_2$(OR) | P-value | Diseases | Evidence |
|---|---|---|---|---|---|
| AHR | Amodiaquine Pathway (PK) | 11.6 | 8.4E-07 | N/A | N/A |
| | Warfarin Pathway (PK) | 9.8 | 7.9E-06 | N/A | N/A |
| | Estrogen Metabolism Pathway | 9.6 | 4.9E-08 | N/A | N/A |
| | Erlotinib Pathway (PK) | 9.6 | 1.0E-05 | N/A | N/A |
| | Phenytoin Pathway (PK) | 8.3 | 5.3E-05 | N/A | N/A |
| ATF1 | Busulfan Pathway (PD) | 4.0 | 6.9E-05 | Cancer | Mutation: melanoma (Futreal et al. 2004) |
| ATF2 | ACE Inhibitor Pathway (PD) | 6.3 | 3.0E-05 | Hypertension | - |
| | Agents Acting on the Renin-Angiotensin System Pathway (PD) | 6.3 | 3.0E-05 | | |
| BRF1 | Imatinib Pathway (PK & PD) | 8.0 | 8.0E-05 | Cancer | - |
| BRF2 | Imatinib Pathway (PK & PD) | 7.7 | 1.2E-04 | Cancer | - |
| CREB1 | Sympathetic Nerve Pathway (Neuroeffector Junction) | 4.2 | 5.4E-06 | Neural normal& | Member: Reactome REACT_18334.1, REACT_15370.2 |
| E2F1 | Antimetabolite Pathway - Folate Cycle (PD) | 4.8 | 2.5E-07 | Cancer & autoimmune diseases | Member: multiple GO terms for Cell cycle control |
| | Thiopurine Pathway (PK & PD) | 3.1 | 3.2E-05 | Cancer & autoimmune diseases | Member: multiple GO terms for Cell cycle control |
| | Methotrexate Pathway (Cancer Cell) (PD) | 3.1 | 1.3E-04 | Cancer & autoimmune diseases | Member: multiple GO terms for Cell cycle control |
| E2F4 | Antimetabolite Pathway - Folate Cycle (PD) | 3.5 | 2.7E-05 | Cancer & autoimmune diseases | Member: multiple GO terms for Cell cycle control |
| EGR1 | EGFR Inhibitor Pathway (PD) | 2.6 | 4.5E-05 | Cancer | - |
| ELK1 | EGFR Inhibitor Pathway (PD) | 5.6 | 5.8E-06 | Cancer | Member: PharmGKB (EGFR Inhibitor Pathway) |
| ETS1 | Vinka Alkaloid Pathway (PK) | 4.9 | 6.2E-05 | N/A | N/A |
| | Doxorubicin Pathway (Cancer Cell) (PD) | 4.1 | 5.2E-05 | Cancer | Member: PharmGKB (EGFR Inhibitor Pathway) |
| | Vemurafenib Pathway (PD) | 3.9 | 2.2E-05 | Cancer (melanoma) | Member: PharmGKB (EGFR Inhibitor Pathway) |
| | Platelet Aggregation Inhibitor Pathway (PD) | 3.2 | 2.0E-05 | thrombosis | - |
| ETV4 | ACE Inhibitor Pathway (PD) | 5.6 | 1.1E-04 | hypertension | - |
| | Agents Acting on the Renin-Angiotensin System Pathway (PD) | 5.6 | 1.1E-04 | | |
| | Celecoxib Pathway (PD) | 4.6 | 8.2E-06 | pain | - |
| FOS | Platinum Pathway (PK & PD) | 4.4 | 3.2E-05 | Cancer | Member: PharmGKB (EGFR Inhibitor Pathway) |
| | Doxorubicin Pathway (Cancer Cell) (PD) | 4.1 | 1.3E-05 | Cancer | Member: PharmGKB (EGFR Inhibitor Pathway) |
| FOXA1 | Benzodiazepine Pathway (PK) | 4.6 | 1.1E-04 | N/A | N/A |
| FOXA2 | Carbamazepine Pathway (PK) | 4.4 | 2.1E-05 | N/A | N/A |
| | Phenytoin Pathway (PK) | 4.0 | 7.0E-05 | N/A | N/A |
| HIF1A | Glucocorticoid Pathway (Peripheral Tissue) (PD) | 6.0 | 6.3E-05 | Inflammation | Member: GO cytokine production |
| HNF1A | Artemisinin and Derivatives Pathway (PK) | 8.4 | 5.4E-07 | N/A | N/A |
| | Tramadol (PK) | 8.1 | 9.4E-07 | N/A | N/A |
| | Losartan Pathway (PK) | 8.0 | 9.2E-05 | N/A | N/A |
| | Sorafenib(PK) | 7.6 | 1.4E-04 | N/A | N/A |
| | Mycophenolic acid Pathway (PK) | 7.4 | 1.2E-09 | N/A | N/A |
| | Irinotecan Pathway (PD*) | 7.3 | 7.4E-08 | Cancer | Mutation: renal cell carcinoma (Rebouissou et al. 2005) |
| | Benzodiazepine Pathway (PK) | 7.3 | 4.4E-06 | N/A | N/A |
| | Valproic Acid Pathway (PK) | 7.0 | 4.0E-09 | N/A | N/A |
| | Irinotecan Pathway (PK) | 7.0 | 7.6E-06 | N/A | N/A |
| | Tamoxifen Pathway (PK) | 6.9 | 2.3E-07 | N/A | N/A |
| | Estrogen Metabolism Pathway | 6.7 | 1.2E-05 | N/A | N/A |
| | Phenytoin Pathway (PK) | 6.3 | 3.0E-05 | N/A | N/A |
| | Statin Pathway (PD) | 5.9 | 5.9E-05 | High cholesterol | Association (Teslovich et al. 2010) |
| | Anti-diabetic Drug Potassium Channel Inhibitors Pathway (PD) | 5.7 | 8.4E-05 | Diabetes | Mutation: diabetes, type 1 (Yamada et al. 1997) and type 2 (Hegele et al. 1999) |
| HNF1B | Methotrexate Pathway (PK) | 9.3 | 1.5E-05 | N/A | N/A |
| HNF4G | Statin Pathway (PD) | 4.7 | 6.5E-05 | High cholesterol | Weak Association (Kathiresan et al. 2007) |
| HOXA5 | Doxorubicin Pathway (Cancer Cell) (PD) | 8.2 | 6.0E-05 | Cancer | Mouse: abnormal cell migration (Mandeville et al. 2006) |
| HSF1 | Glucocorticoid Pathway (Peripheral Tissue) (PD) | 7.1 | 6.4E-06 | Inflammation | Mouse (Xiao et al. 1999) |
| KLF11 | Anti-diabetic Drug Potassium Channel Inhibitors Pathway (PD) | 9.2 | 2.1E-05 | Diabetes | Mutation (Neve et al. 2005) |
| MYC | Antimetabolite Pathway - Folate Cycle (PD) | 3.9 | 4.5E-06 | Cancer & autoimmune diseases | Member: GO cell cycle control; Census (Collins and Groudine 1982; Yokota et al. 1986); |
| | Thiopurine Pathway (PK & PD) | 2.4 | 8.6E-05 | Cancer & autoimmune diseases | Member: GO cell cycle control; Census (Collins and Groudine 1982; Yokota et al. 1986); |
| NFKB1 | Doxorubicin Pathway (Cancer Cell) (PD) | 4.4 | 3.8E-06 | Cancer | - (Strong literature evidence but no mutation) |
| | EGFR Inhibitor Pathway (PD) | 3.1 | 8.9E-07 | Cancer | - |
| NR1I2 | Clopidogrel Pathway (PK) | 6.1 | 4.7E-05 | N/A | N/A |
| NR1I3 | Fluvastatin Pathway (PK) | 11.3 | 1.5E-06 | N/A | N/A |
| | Atorvastatin/Lovastatin/Simvastatin Pathway (PK) | 11.2 | 1.7E-06 | N/A | N/A |
| | Statin Pathway - Generalized(PK) | 10.7 | 3.6E-06 | N/A | N/A |
| | Phenytoin Pathway (PK) | 10.7 | 3.6E-06 | N/A | N/A |
| NR2C2 | Imatinib Pathway (PK & PD) | 7.7 | 1.2E-04 | Cancer | - |
| PDX1 | Anti-diabetic Drug Potassium Channel Inhibitors Pathway (PD) | 9.2 | 2.1E-05 | Diabetes | Mutation (Macfarlane et al. 1999; Hani et al. 1999) |
| PHOX2A | Sympathetic Nerve Pathway (Neuroeffector Junction) | 9.9 | 9.9E-06 | Neural normal[&] | Mouse (Coppola et al. 2010) |
| PPARA | Statin Pathway (PD) | 6.3 | 3.8E-08 | High cholesterol | Mutation (Vohl et al. 2000) |
| | Celecoxib Pathway (PD) | 4.6 | 7.8E-05 | Pain | - |
| PPARD | Statin Pathway (PD) | 6.4 | 2.1E-05 | High cholesterol | Mouse (Gross et al. 2011) |
| RARA | Aromatase Inhibitor Pathway (Breast Cell) (PD) | 8.1 | 1.6E-06 | Cancer | Census |
| | EGFR Inhibitor Pathway (PD) | 4.0 | 1.1E-05 | Cancer | Census |
| RARB | Vemurafenib Pathway (PD) | 5.6 | 1.0E-04 | Cancer | - |
| | EGFR Inhibitor Pathway (PD) | 4.5 | 1.0E-04 | Cancer | - |
| RDBP | Imatinib Pathway (PK & PD) | 8.0 | 8.0E-05 | Cancer | - (Involved in cancer, not through genetic mutation) |
| REST | Sympathetic Nerve Pathway (Pre- and Post- Ganglionic Junction) | 4.9 | 6.9E-05 | Neural normal[&] | Mouse: thick retinal ganglion layer (Mao et al. 2011) |
| RXRA | Statin Pathway (PD) | 5.4 | 6.1E-08 | High cholesterol | - |
| SP1 | Vinka Alkaloid Pathway (PK) | 4.6 | 5.2E-06 | N/A | N/A |
| | Erlotinib Pathway (PK) | 4.3 | 5.7E-05 | N/A | N/A |
| | Etoposide Pathway (PK & PD) | 3.9 | 7.3E-06 | Cancer | - |
| | Statin Pathway (PD) | 3.8 | 3.4E-08 | High cholesterol | - |
| | EGFR Inhibitor Pathway (PD) | 2.1 | 7.3E-05 | Cancer | - |
| SREBF1 | Bisphosphonate Pathway (PD) | 5.7 | 3.0E-07 | Osteoporosis | - |
| | Statin Pathway (PD) | 5.3 | 7.0E-08 | High cholesterol | Mouse Knockout (Liang et al. 2002) |
| SREBF2 | Bisphosphonate Pathway (PD) | 6.8 | 9.3E-09 | Osteoporosis | - |
| | Statin Pathway (PD) | 6.4 | 1.1E-09 | High cholesterol | Mouse Knockout (Liang et al. 2002) |
| STAT1 | EGFR Inhibitor Pathway (PD) | 3.1 | 1.9E-05 | Cancer | - |
| STAT5A | Aromatase Inhibitor Pathway (Breast Cell) (PD) | 8.9 | 2.8E-05 | Breast & ovarian cancer | - |
| STAT5B | Aromatase Inhibitor Pathway (Breast Cell) (PD) | 8.9 | 2.6E-05 | Breast & ovarian cancer | Cancer Gene Census |
| TFAP2A | Sympathetic Nerve Pathway (Neuroeffector Junction) | 3.6 | 2.4E-06 | Neural normal[&] | Mutation: Branchio-oculo-facial syndrome (Lin et al. 2000; Milunsky et al. 2008; Gestri et al. 2009) |
| | Celecoxib Pathway (PD) | 3.2 | 4.3E-07 | Pain | |
| | EGFR Inhibitor Pathway (PD) | 2.6 | 8.8E-05 | Cancer | - |
| TP53 | Doxorubicin Pathway (Cancer Cell) (PD) | 5.2 | 2.0E-09 | Cancer | Census (Chen et al. 1990; Halevy et al. 1990) |
| | Vinka Alkaloid Pathway (PK) | 5.0 | 4.6E-05 | - | - |
| | Etoposide Pathway (PK & PD) | 4.6 | 1.3E-05 | Cancer | Census (Chen et al. 1990; Halevy et al. 1990) |
| | Busulfan Pathway (PD) | 4.2 | 3.9E-09 | Cancer | Census (Chen et al. 1990; Halevy et al. 1990) |
| | Vemurafenib Pathway (PD) | 4.0 | 1.5E-05 | Cancer | Census (Chen et al. 1990; Halevy et al. 1990) |
| | Methotrexate Pathway (Cancer Cell) (PD) | 4.0 | 2.6E-06 | Cancer & autoimmune diseases | Census (Chen et al. 1990; Halevy et al. 1990); Normal immune response (Chiang et al. 2012) |
| | Doxorubicin Pathway (PK) | 3.9 | 1.0E-04 | - | - |
| | EGFR Inhibitor Pathway (PD) | 3.4 | 2.4E-08 | Cancer | Census (Chen et al. 1990; Halevy et al. 1990) |
| USF1 | Theophylline Pathway (PK) | 6.0 | 1.2E-04 | - | - |
| WT1 | Doxorubicin Pathway (Cancer Cell) (PD) | 6.6 | 4.6E-07 | Cancer | Census (Pelletier et al. 1991) |
| | Doxorubicin Pathway (PK) | 5.8 | 6.7E-05 | - | - |
| | Vemurafenib Pathway (PD) | 5.6 | 1.0E-04 | Cancer | Census (Pelletier et al. 1991) |
| YBX1 | Erlotinib Pathway (PK) | 9.9 | 6.7E-06 | - | - |

**Table S5.** The tables list the functional concepts with the highest and lowest unique scores from 6 sources of gene function annotations. The uniqueness score are computed with all data sources merged.

| GWAS Phenotypes | Uniqueness | Mendelian Diseases | Uniqueness |
|---|---|---|---|
| oligospermia | 1 | anauxetic dysplasia | 1 |
| cd8-positive t-lymphocytes | 1 | polycystic ovary syndrome | 1 |
| t-lymphocytes | 1 | intracranial arterial disease | 1 |
| serotonin | 1 | Moyamoya disease | 1 |
| transforming growth factor beta1 | 1 | Mobius syndrome | 1 |
| lymphoma, follicular | 0.24 | papilloma | 0.1 |
| hepatitis b, chronic | 0.24 | choroid plexus papilloma | 0.1 |
| myeloproliferative disorders | 0.23 | adrenocortical carcinoma | 0.1 |
| alopecia | 0.22 | adrenal gland cancer | 0.1 |
| papillomaviridae | 0.18 | adrenal cortex cancer | 0.1 |

| PharmGKB PK/PD pathways | Uniqueness | GO MF | Uniqueness |
|---|---|---|---|
| PA165947317_Leukotriene_modifiers_pathway_PD | 0.75 | DRUG_BINDING | 0.76 |
| PA2026_Glucocorticoid_Pathway_(HPA_Axis)_PD | 0.73 | PROTEIN_TRANSPORTER_ACTIVITY | 0.74 |
| PA165111376_Benzodiazepine_Pathway_PD | 0.66 | OXIDOREDUCTASE_ACTIVITY_ACTING_ON_SULFUR_GROUP_OF_DONORS | 0.72 |
| PA2027_Glucocorticoid_Pathway_(Peripheral_Tissue)_P | 0.65 | OXIDOREDUCTASE_ACTIVITY_ACTING_ON_THE_CH_N | 0.7 |
| D | | H_GROUP_OF_DONORS | |
| PA2036_Gemcitabine_Pathway_PD | 0.59 | NUCLEOBASENUCLEOSIDENUCLEOTIDE_AND_NUCLEIC_ACID_TRANSMEMBRANE_TRANSPORTER_ACTIVITY | 0.68 |
| PA145011111_Fluvastatin_Pathway_PK | 0.11 | METAL_ION_TRANSMEMBRANE_TRANSPORTER_ACTIVITY | 0.08 |
| PA152325160_Gefitinib_Pathway_PK | 0.1 | ION_CHANNEL_ACTIVITY | 0.08 |
| PA2034_Cyclophosphamide_Pathway_PK | 0.1 | TRANSMEMBRANE_TRANSPORTER_ACTIVITY | 0.08 |
| PA145011108_Statin_Pathway_-_Generalized_PK | 0.1 | ION_TRANSMEMBRANE_TRANSPORTER_ACTIVITY | 0.07 |
| PA145011109_Atorvastatin/Lovastatin/Simvastatin_Pathway_PK | 0.09 | SUBSTRATE_SPECIFIC_TRANSMEMBRANE_TRANSPORTER_ACTIVITY | 0.07 |

| GO BP | Uniqueness | Signaling/Metabolic Pathways | Uniqueness |
|---|---|---|---|
| ESTABLISHMENT_AND_OR_MAINTENANCE_OF_CELL_POLARITY | 0.63 | TRYPTOPHAN_CATABOLISM | 0.86 |
| KERATINOCYTE_DIFFERENTIATION | 0.58 | DIGESTION_OF_DIETARY_CARBOHYDRATE | 0.76 |
| RESPIRATORY_GASEOUS_EXCHANGE | 0.58 | VITAMIN_B5_PANTOTHENATE_METABOLISM | 0.72 |
| PEPTIDE_METABOLIC_PROCESS | 0.58 | METABOLISM_OF_POLYAMINES | 0.72 |
| MEMBRANE_FUSION | 0.58 | CELL_EXTRACELLULAR_MATRIX_INTERACTIONS | 0.7 |
| REGULATION_OF_TRANSCRIPTION | 0.06 | CDT1_ASSOCIATION_WITH_THE_CDC6_ORC_ORIGIN_COMPLEX | 0.04 |
| REGULATION_OF_NUCLEOBASENUCLEOSIDENUCLEOTIDE_AND_NUCLEIC_ACID_METABOLIC_PROCESS | 0.06 | P53_INDEPENDENT_G1_S_DNA_DAMAGE_CHECKPOINT | 0.04 |
| REGULATION_OF_GENE_EXPRESSION | 0.06 | SCF_BETA_TRCP_MEDIATED_DEGRADATION_OF_EMI1 | 0.04 |
| REGULATION_OF_METABOLIC_PROCESS | 0.06 | AUTODEGRADATION_OF_THE_E3_UBIQUITIN_LIGASE_COP1 | 0.04 |
| REGULATION_OF_CELLULAR_METABOLIC_PROCESS | 0.06 | CDK_MEDIATED_PHOSPHORYLATION_AND_REMOVAL_OF_CDC6 | 0.04 |

**Table S6:** A list of negatively associated functional concepts linked by transcription factors.
**File:** TableS6.xlsx

**Table S7**: The complete TF annotation results. -$Log_{10}$ (p-value) are provided in parentheses following the target functions.

**File:** TableS7.xlsx